\documentclass[12pt]{iopart}

\usepackage[english]{babel} 
\usepackage[utf8]{inputenc}
\usepackage{color}
\usepackage{graphicx}
\renewcommand{\vec}[1]{\mathbf{#1}}

\begin{document}

\title[RG flow of a weak extended backscattering Hamiltonian in a non-chiral TLL]{Renormalization flow of a weak extended backscattering Hamiltonian in a non-chiral Tomonaga-Luttinger liquid}

\author{A. Popoff$^{1,2}$, A.V. Lebedev$^{3}$ , L. Raymond$^{1}$, T. Jonckheere$^{1}$, J. Rech$^{1}$, T. Martin$^{1}$}

\address{$^{1}$ Aix Marseille Univ, Université de Toulon, CNRS, CPT, Marseille, France}
\address{$^{2}$ Collège Tinomana Ebb de Teva I Uta, BP 15001 - 98726 Mataiea, Tahiti, French Polynesia}
\address{$^{3}$ Moscow Institute of Physics and Technology, Institutskii per. 9, Dolgoprudny, 141700, Moscow District, Russia}

%\ead{@cpt.univ-mrs.fr}

\begin{abstract}
We consider a non-chiral Luttinger liquid in the presence of a backscattering Hamiltonian which has an extended range. Right/left moving fermions at a given location can thus be converted as left/right moving fermions at a different location, within a specific range. We perform a momentum shell renormalization group treatment which gives the evolution of the relative degrees of freedom of this Hamiltonian contribution under the renormalization flow, and we study a few realistic examples of this extended backscattering Hamiltonian. We find that, for repulsive Coulomb interaction in the Luttinger liquid, \emph{any} such Hamiltonian contribution evolves into a delta-like scalar potential upon renormalization to a zero temperature cutoff. On the opposite, for attractive couplings, the amplitude of this kinetic Hamiltonian is suppressed, rendering the junction fully transparent. As the renormalization procedure may have to be stopped because of experimental constraints such as finite temperature, we predict the actual spatial shape of the kinetic Hamiltonian at different stages of the renormalization procedure, as a function of the position and the Luttinger interaction parameter, and show that it undergoes structural changes. This renormalized kinetic Hamiltonian has thus to be used as an input for the perturbative calculation of the current, for which we provide analytic expressions in imaginary time. We  discuss the experimental relevance of this work by looking at one-dimensional systems consisting of carbon nanotubes or semiconductor nanowires.  
\end{abstract}


\submitto{\JPCM}

%%%%%%%%%%%%%%%%%%%
\section{Introduction}

Luttinger liquids~\cite{luttinger_65,mattis_65,haldane_81} (LLs) constitute an important paradigm of theoretical condensed matter physics. In the context of quantum nanophysics, they are very good candidates for explaining the transport properties of correlated one-dimensional systems. Many techniques for the fabrication of nanowires with large mobility are now available, and naturally occurring one-dimensional systems such as carbon nanotubes are routinely employed in electronic quantum transport (all constitute  strong candidates for LL), generating a dialog between experimentalists and theorists. 

In quantum transport, LLs made their appearance in the early nineties. Several authors~\cite{kane_92a,kane_92b,kane_92c,furusaki_93a} established the phase diagram of the transport properties of a one-dimensional wire with a delta function impurity, as a function of the interaction parameter $g$ of the LL and the transmission of the barrier. Indeed, it was noticed that ~\cite{kane_92a,kane_92b,kane_92c,furusaki_93a} when integrating out the quadratic bosonic degrees of freedom of the Luttinger Liquid Euclidean action away from the impurity location, the resulting action with fields evaluated at the location of the impurity corresponds to that of a particle in a periodic potential coupled to a bath of harmonic oscillators~\cite{guinea_85,fisher_85}. These results allowed to establish a one-to-one mapping between the two systems, and in particular their phase diagram. 

In the presence of backscattering, a barrier with high or low transmission renders the system insulating when Coulomb interactions ($g<1$) are operating, while in the opposite case of attractive interactions ($g>1$), the barrier is transparent.  In the dual situation, where effectively two semi-infinite wires communicate by a tunnel hopping amplitude, Coulomb interactions ($g<1$) lead as expected to an insulating behavior, while attractive interactions ($g>1$) result in perfect transmission. These results were further extended to treat resonant tunneling in double and periodic barriers ~\cite{kane_92b,kane_92c,furusaki_93b}. In the case of a double barrier, it was shown that there is a possibility for electrons subject to Coulomb repulsion to be nevertheless transmitted when  $1/2<g<1$, an effect known as resonant tunneling. It was further argued ~\cite{martin_95,sanjose_05} that when coupling a LL to acoustic phonons, the phase diagram is substantially modified by the attractive retardation effects mediated by phonons. Quantum impurity problems in the context of LLs is still an active field of research, with recent results demonstrating the equivalence between two distinct LL impurity problems.~\cite{kane_20}

LLs are also very useful for their predictive power in the transport properties of the fractional quantum Hall effect (FQHE). Indeed, edge excitations occur at the boundaries of the FQHE bar, described by the chiral version of the LL. This allowed~\cite{kane_94,chamon_95} to characterize the transport properties of a Hall bar separated by a quantum point contact (QPC). In turn, these results lead to an experimental proposal allowing to identify the fractional charge of Laughlin quasiparticles by measuring the backscattering current and noise.~\cite{depicciotto_97,saminadayar_97}

The renormalization group  (RG) is a very powerful tool for most fields of low and high energy physics (for a presentation in the context of condensed matter physics, see \cite{chaikin,altland}). Here in low dimensional quantum transport, it allows to derive phase diagrams as a function of the LL interaction parameter. In its perturbative version (integrating momentum/frequency degrees of freedom on a shell close to the Fermi energy) it enables to specify the evolution of the coupling constants of the system under renormalization. At zero temperature, one needs to push the renormalization until convergence is ultimately reached. However, since the experiment is carried out at a finite temperature $T_0$, the renormalization procedure should be stopped when a lower cutoff is reached, such as the lowest Matsubara frequency $2\pi k_B T_0$. Other cutoffs may dictate that the renormalization of the coupling parameters should be stopped, e.g. when a specific (external) frequency scale associated with a perturbation is reached, or when the frequency related to the length scale of the system $L$ ($\omega_L= \pi u/L$, $u$ Fermi velocity) is attained. The (perturbative) RG approach is therefore not only relevant for establishing the ultimate convergence point, but also for identifying the coupling constants when the procedure has to be stopped because of the experimental context.

The analysis of impurity/tunneling effects in both non-chiral and chiral LLs have been achieved mostly so far assuming that the backscattering/tunneling location is point-like as described by a scalar potential. Yet in practice, the presence of an impurity generates a kinetic backscattering Hamiltonian with a  finite extent, and the situation is even more complex if several impurities/tunneling locations are present. This is relevant for both non-chiral LL which describe Coulomb interactions between electrons in the nanowire as well as chiral LL where the edge excitations describe fractional quasiparticles in the quantum Hall effect.  When a QPC pinches off a FQHE bar, this QPC is placed at distance above the two-dimensional electron gas, and the tunneling region between counter-propagating edge states should also have a finite range. This situation has been only studied in some specific instances so far. In the FQHE, backscattering in an  extended tunneling region has been addressed,~\cite{chevallier_10b} with the informative result that the Fano factor associated with Laughlin quasiparticle is unchanged, in the Poissonian limit (lowest order in the tunneling amplitude). 

\begin{figure}[tbp]
	\centering
		\includegraphics[width=0.7\textwidth]{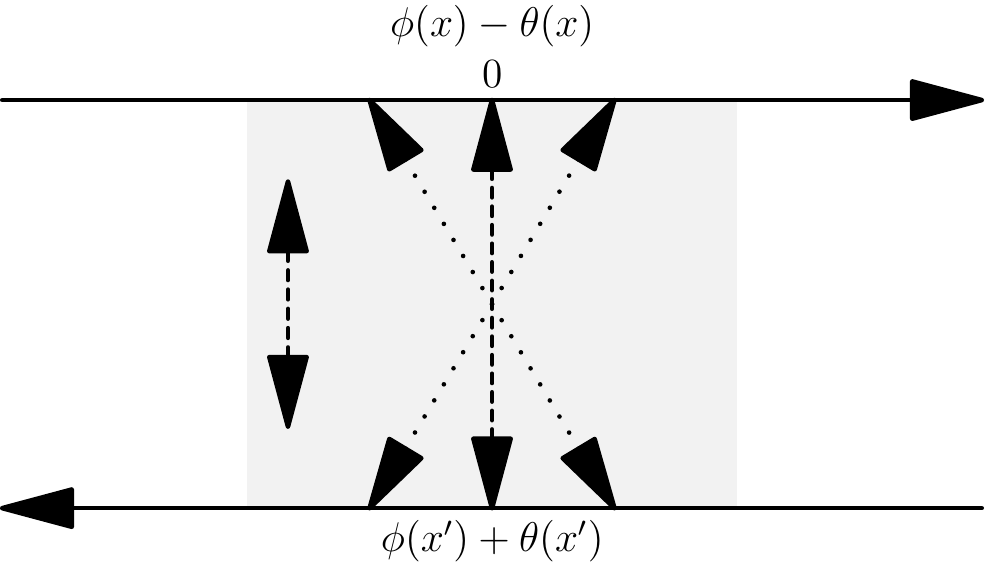}
		\caption{Schematic representation of a one-dimensional wire with an extended backscattering Hamiltonian. Right and left moving fermions are depicted by full horizontal lines, and the phases of the right/left fermion fields appear as $\theta(x)\mp \phi(x)$. Vertical dashed lines with arrows indicate a backscattering path which can be described by a scalar potential as in~\cite{aranzana_05}. Oblique tunneling paths correspond to the finite range processes which delocalize right and left moving electrons, and constitute the focus of this study. The gray area depicts the region where the amplitude of the extended backscattering Hamiltonian, centered at $x=0$, is appreciable.}
		\label{fig00}
\end{figure}

Extended backscattering Hamiltonians (XBHs) are likely to be present in one-dimensional physical systems which are relevant for experiments. For metallic carbon nanotubes, which constitute strong candidates for multi-mode LLs, this backscattering can occur because of the presence of impurities/defects located in a finite region of the wire, or alternatively because such nanotubes contain a ``cusp'' or a ``bend''.~\cite{farajian_03} In the latter case, the extent of the backscattering Hamiltonian region is typically proportional to the radius of curvature of this bent region. Semiconductor nano-wires allow more flexibility, as metallic gates  placed in  the vicinity of the nanowire can be tailored to generate an ``extended impurity'' region. Right (left) moving electrons can  be for instance backscattered into left (right) moving electrons at a specific position over a finite range. This case of a purely local backscattering potential (generalized with a finite extent) has been discussed in \cite{aranzana_05},  and involve only minor modifications/adjustments of the formalism of~\cite{kane_92a,kane_92b,kane_92c}: this mechanism is represented by the ``vertical'' paths for electron tunneling in Fig.~\ref{fig00}. Alternatively a right (left) moving electron can be destroyed at a specific location and be transferred as a left (right) moving electron at  a {\it different} location (a process described by an oblique tunneling path in Fig.~\ref{fig00}). In a tight binding picture, such processes are equivalent to next near neighbor hopping and so on. Contrary to local tunneling they are described by a kinetic contribution to the Hamiltonian rather than a scalar potential as in~\cite{aranzana_05}. To our knowledge, such an XBH has not been considered so far in the literature, and no analysis of its renormalization flow has been made available in the context of LLs. 

In the present paper, we intend to bridge this gap, for the case of non-chiral LLs. Our aim is therefore to study the effect of weak backscattering due to an interaction of arbitrary shape, which includes a scalar potential as well as a hopping (kinetic) interaction over some region of space. We introduce the LL model with this XBH in Sec.~\ref{LL_model}. We derive the RG equation for the latter in Sec.~\ref{momentum_shell}, focusing on a kinetic Hamiltonian with a finite range and centered at a specific location. This, together with the two applications described in this work,  constitute the central results of this paper. We illustrate our renormalization results in Sec.~\ref{relative_potential_section} with numerics showing the evolution of the RG procedure for both an extended localized interaction and a double barrier/extended interaction. One (expected) result that we recover is that if one pushes the renormalization procedure to zero temperature, {\it any} XBH is washed away when interactions are attractive ($g>1$), but with repulsive Coulomb coupling ($g<1$), it converges to the stable fixed point
which corresponds to a delta-like function (scalar) potential. More importantly, along the way of the RG flow (when renormalization may have to be stopped because of experimental constraints) the initial XBH undergoes major shape changes where new length scales appear in this kinetic Hamiltonian profile. In Sec.~\ref{perturbative_current}, we derive the linear response theory for the backscattering current associated with such XBH and show how it collapses to the result where the potential is purely local.~\cite{kane_92b,kane_92c} A discussion about the applicability and relevance of our results in physical situations is presented in Sec. \ref{discussion}. We conclude in Sec.~\ref{conclusion}. 

%%%%%%%%%%%%%%%%%%%%%%%%%%%%%%%%%%%%%%%%%%%%%%%%%%%%%%%%%%%%%%%%%%%%%%%

\section{Model}\label{LL_model}

\subsection{Luttinger model and backscattering interaction}

We consider an (infinite) one-dimensional interacting electron nanowire in the presence of an XBH.
We focus for simplicity on spinless fermions (i.e. in the absence of spin flip processes, since adding the spin degrees of freedom is a mere formality\cite{kane_92b,kane_92c}). The fermion annihilation operators are then a superposition of fields emanating from the right and left:
\begin{eqnarray}
\psi(x)&=\frac{1}{\sqrt{\eta}} \left[ e^{ik_Fx}e^{i\sqrt{\pi} \left( \phi(x)-\theta(x) \right)}  + e^{-ik_Fx}e^{i\sqrt{\pi} \left( \phi(x)+\theta(x) \right) }\right],
\label{eq:psi}
\end{eqnarray}
where  $\eta$ is the short wave length cutoff of the Luttinger model, while $\theta(x)$ and $\phi(x)$ are the bosonic fields of the non-chiral LL model, which satisfy the canonical commutation relation:
\begin{eqnarray}
\left[ \phi(x),\theta(x')\right]=-\frac{i}{2}\mathrm{sgn}(x-x').
\label{relcano}
\end{eqnarray}
The free Hamiltonian reads: 
\begin{eqnarray}
H_0=\frac{u}{2}\int dx\,\left[g(\partial_x\phi)^2+g^{-1}(\partial_x\theta)^2\right],
\label{bare}
\end{eqnarray}
where $u$ is proportional to the bare Fermi velocity and $g$ is the LL interaction parameter, with $g<1$ for repulsive (Coulomb) interaction, $g=1$ for non-interacting fermions, and $g>1$ for attractive interactions (in what follows, we set $u=1$ for simplicity).
The interaction responsible for backscattering effects has then the general form:
\begin{eqnarray}
H_{\rm{imp}}=\int dx\,dx'\,\bigl[ V(x,x')\psi^\dagger(x)\psi(x')+ \rm{H.c.}\bigr].
\label{backscattering}
\end{eqnarray}
If $V(x,x')\sim \delta(x-x')$, this interaction corresponds to a scalar potential, but otherwise, it describes a kinetic hopping term where right (left) moving fermions at a given location are converted into left (right) moving fermions at a different locations, as depicted in Fig.~\ref{fig00}. 
In terms of the bosonic fields, the contribution of Eq.~(\ref{backscattering}) which describes solely backscattering effects reads:
\begin{eqnarray}
H_{\rm{imp}}'=\int dx & \,dx'\, \eta^{-1}  V(x,x') \nonumber \\
 & \times \cos\left\{ \sqrt \pi\bigl[ \theta(x)+\theta(x')+\phi(x)-\phi(x')\bigr] +k_F(x+x') \right\},
\label{HRD}
\end{eqnarray}
while other contributions  in Eq.~(\ref{backscattering}) are linear in the bosonic fields, and can be reabsorbed in the bare action resulting from Eq.~(\ref{bare}) giving rise to a mere modification of the LL parameters $u$ and $g$.~\cite{kane_92a,kane_92b,kane_92c} 

\subsection{Euclidean action and Green's functions}

The partition function of the system is then expressed as a functional integral over the bosonic fields $\theta$ and $\phi$:
\begin{eqnarray}
Z=\int \mathcal D\theta\, \mathcal D\phi\, e^{-S(\theta, \phi)},
\end{eqnarray}
where $S=S_0+S_{\rm{imp}}$ is the total action, and $S_0, S_{\rm{imp}}$ are respectively the free Euclidean action and the action associated with the XBH.

The free Euclidean action in imaginary time reads:
\begin{eqnarray}
      S_0=\frac12\int_0^\beta d\tau\,\int dx\,\Bigl[ g(\partial_x
      \phi)^2+g^{-1}(\partial_x\theta)^2-2i\partial_\tau
      \phi\partial_x\theta\Bigr],
      \label{aceuclid}
\end{eqnarray}
which is conveniently written in Fourier space
\begin{eqnarray}
      S_0&=\frac12\sum_{\vec p}\left[ q^{2}g \phi^*_{\vec p}\phi_{\vec p}+\frac{q^2}g \theta^*_{\vec p}\theta_{\vec p}-i\omega q \phi^*_{\vec p}\theta_{\vec p} -i\omega q\theta^*_{\vec p}\phi_{\vec p}\right],
      \label{bare_action_fourier}
\end{eqnarray}
with $\vec p=(q,\omega)$. For subsequent calculations (for instance when performing the average over the fast degrees of freedom of the impurity action), it is useful to introduce the following bare Green's functions (in their Fourier transform version):
\begin{eqnarray}
\left(
         \begin{array}{cc}
 G^{\phi\phi}(\vec{p}) & G^{\phi\theta}(\vec{p}) \\
 G^{\theta\phi}(\vec{p}) & G^{\theta\theta}(\vec{p}) 
        \end{array}
    \right)
	=\frac{1}{\vec{p}^2}
\left(
         \begin{array}{cc}
	g^{-1} & i\omega/q \\
	i\omega/q & g 
        \end{array}
    \right),
\label{fourier_green}
\end{eqnarray}
while the Green's functions in space/imaginary time read:
\begin{eqnarray}
      G^{\phi\phi}(z,\tau)=\int\limits_{|\vec p|\leq\Lambda} \frac{d^2 p}{(2\pi)^2}\,G^{\phi\phi}(\vec p) e^{iqz+i\omega \tau},
\end{eqnarray}
where $\Lambda$ is a large momentum/frequency cutoff, and similar expressions are defined for $\theta\phi$, $\phi\theta$, $\theta\theta$ Green's functions. Note that for a purely local backscattering potential or a weak link separating two semi-infinite LLs~\cite{kane_92a,kane_92b,kane_92c,furusaki_93b}, only the diagonal elements $G^{\phi\phi}$ and $G^{\theta\theta}$ are needed. Off diagonal elements $G^{\phi\theta}$ and $G^{\theta\phi}$ are for instance unavoidable in transport problems involving the injection of electrons in the bulk of LLs.~\cite{crepieux_03,lebedev_95,guigou_07,guigou_09a,guigou_09b} In the present ``extended'' impurity problem, all Green's functions are needed because the impurity action contains both fields $\theta$ and $\phi$ [see Eq.~(\ref{impurity_xg_z}) below].

The contribution to the action associated with the XBH is more conveniently described after the following change of coordinates is performed on the interaction potential: 
\begin{eqnarray}
V(x,x^\prime)=\bar{V}(x_G,z),
\end{eqnarray}
where $x_G=(x+x')/2$ is the center of mass coordinate and $z=x-x'$ is the relative coordinate. Concerning the dependence of  $\bar{V}(x_G,z)$ on the center of mass coordinate $x_G$, we assume that it is restricted to a finite range characterized by a length scale $L$, while the dependence on the relative coordinate $z$ occurs within a range $a$. Note that in general, both $L$ and $a$ should be chosen to be much larger than $\eta$ or (equivalently) the inverse momentum cutoff $\Lambda^{-1}$. $\bar{V}(x_G,z)$, for dimensionality reasons, also contains prefactors with powers of $L$ and $a$. With these notations, the backscattering action [associated with the Hamiltonian in Eq.~(\ref{HRD})] then reads:
\begin{eqnarray}
      S_{\rm{imp}}=\int_0^\beta & d\tau\, \int dx_G\,
      dz\, \eta^{-1}\bar{V}(x_G,z)  \nonumber \\
     &\times \cos \left\{ \sqrt\pi \left[ \theta \left( x_G+ \frac{z}{2},\tau \right) +\theta\left(x_G- \frac{z}{2},\tau\right) \right. \right. \nonumber \\
      & \qquad \left. \left. +\phi\left(x_G+ \frac{z}{2},\tau\right)-\phi\left(x_G- \frac{z}{2},\tau\right) \right]+2k_Fx_G \right\}.
\label{impurity_xg_z}
\end{eqnarray}

%%%%%%%%%%%%%%%%%%%%%%%%%%%%%%%%%%%%%%%%%%%%%%%%%%%%%%%%%%%%%%%%%%%%%%%%

\section{Momentum shell renormalization} \label{momentum_shell}

In this section, we use a perturbative RG approach to deal with the backscattering action $S_{\rm{imp}}$ describing an XBH. 
Let us first recall the basics of the perturbative RG procedure: 
\begin{itemize}
	\item In the total action $S=S_0+S_{\rm{imp}}$, one first identifies the fast and slow components of the fields $\theta$ and $\phi$. This means identifying the momentum/frequency components of these fields which belong to the interval $[\Lambda/b,\Lambda]$ (fast modes), as well as those who belong to the interval $[0,\Lambda/b]$ (slow modes) where $\Lambda$ is the upper cutoff and $b>1$ ($b=1+\epsilon$, $0<\epsilon\ll 1$). One then treats the backscattering interaction to lowest non-vanishing order in perturbation theory, and integrates over the fast degrees of freedom. Upon re-exponentiating, this generates a new action. If this new action has the same form as the preceding one, the theory is called "renormalizable" and one can follow with the next steps.
	
	\item The cutoff of this new action is now  $\Lambda/b$. One then needs to rescale the units of all parameters (length or frequency scales) to reimpose the cutoff $\Lambda$. 

	\item One then obtains a similar action to that of our starting point, with parameters that have changed/evolved under the renormalization procedure. This then allows to derive a differential equation characterizing the evolution of these parameters under the renormalization procedure. Note that in the present situation, there is no specific coupling constant which is renormalized (as in the delta function scalar potential case): rather, the whole XBH amplitude $\bar{V}(x_G,z)$ undergoes modifications upon renormalization. 
\end{itemize}
The details of the renormalization procedure are described in \ref{detailed_RG}. The main results are that first, $\bar{V}(x_G,z)$ is not modified as far as the center of mass coordinate $x_G$ is concerned (because of the translational invariance of the free system), all modifications occur in the dependence with respect to the relative coordinate $z$:
\begin{eqnarray}
 \frac{1}{ \bar{V}(x_G,z,t)}\frac{d \bar{V}}{dt}(x_G,z,t)= \alpha(g,\Lambda,z), 
 \label{Gell-Man_final}
\end{eqnarray}
with 
\begin{eqnarray}
 \alpha(g,\Lambda,z)= 1-g^{-1}+\frac{g^{-1}-g}{2}\bigl[1+J_0(\Lambda |z|)\bigr]\label{alpha},
 \label{eq:alpha}
\end{eqnarray}
where $J_0$ is the Bessel function of zeroth order. Here, $t=\log(b)$ is a fictitious time which describes the evolution under renormalization.
This, together with its applications described in the following section, constitutes the central result of this work.
It suggests that the entire impurity backscattering interaction $\bar{V}(x_G,z)$ is modified as a whole, as illustrated by the presence of the relative spatial coordinate $z$ in the argument of the Bessel function in Eq.~(\ref{alpha}). This is indeed in sharp contrast with previous RG treatments applied to LLs which only lead to a modification of the coupling constants of the initial action. 

\begin{figure}[tbp]
	\centering
		\includegraphics[width=0.7\textwidth]{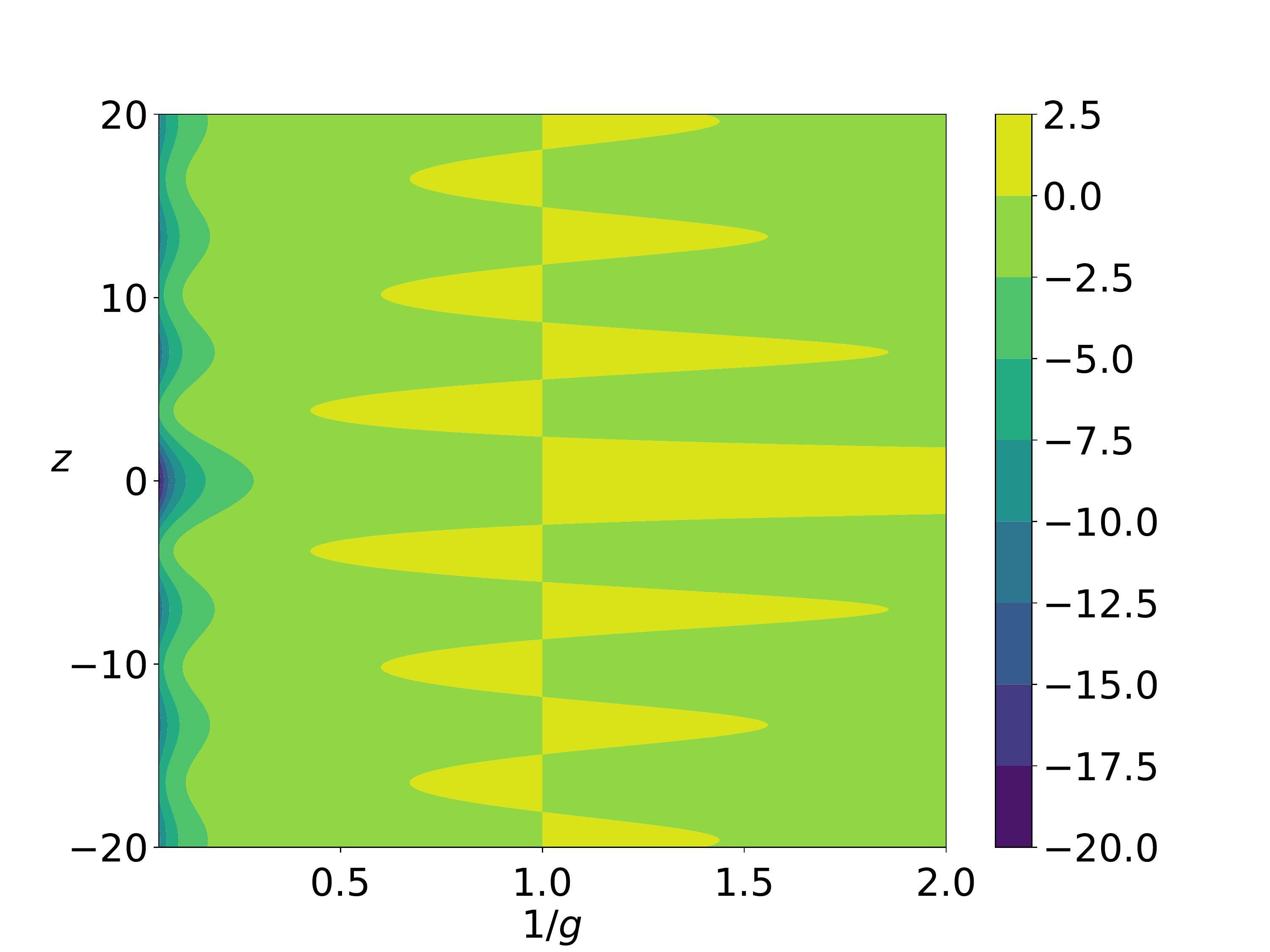}
		\caption{Plot of  $\alpha(g,\Lambda,z)$ as a function of the inverse interaction parameter $1/g$ and the relative coordinate $z$ (in units of $\Lambda^{-1}$). }
		\label{graph_renormalization}
\end{figure}

The backscattering interaction evolves under the RG procedure as:
\begin{eqnarray}
\bar{V}(x_G,z,t)=\bar{V}(x_G,z,0) e^{\int_0^t \alpha(g,\Lambda,z)dt'}.
\end{eqnarray}
It is therefore quite informative to plot the function $\alpha(g,\Lambda,z)$ as a function of the interaction parameter $g$ and the spatial coordinate $z$, which we report in Fig \ref{graph_renormalization}. When  $\alpha(g,\Lambda,z)>0$, the XBH amplitude tends to increase, while it is exponentially suppressed when $\alpha(g,\Lambda,z)<0$. It is thus useful to identify in Fig.~\ref{graph_renormalization} which parameter regions in the $( 1/g , z )$-plane correspond to  positive or negative $\alpha$. One notices oscillations near the $g=1$ line as a function of $z$ which are associated with the zeros of the Bessel function.  These oscillations persist both on the attractive ($g>1$) and the repulsive ($g<1$) sides of Fig.~\ref{graph_renormalization}. For strong attractive interactions of the LL ($g\gg 1$), the backscattering interaction is always suppressed (as illustrated by the dark regions on the left side of the figure), while for strong repulsive interactions ($g\ll 1$) in the LL, one observes a central peak (light colored region on the right side of the figure) which is centered at $z=0$. Although we need to illustrate our results with more specific forms of the interaction potential, we can already make an important statement: when interactions are repulsive, \emph{any} non-zero interaction $\bar{V}(x_G,z)$ evolves into a narrow potential centered at $z=0$, whose amplitude becomes larger when the repulsion is stronger.  On the opposite, in the presence of attractive interactions, the amplitude $\bar V(x_G,z,t)$ of the XBH tends to be reduced under the RG flow.

\section{Application to an extended relative interaction} \label{relative_potential_section}

\subsection{Extended Gaussian potential} \label{extended_gaussian}

\begin{figure*}[tbp]
	\centering
		\includegraphics[width=1.0\textwidth]{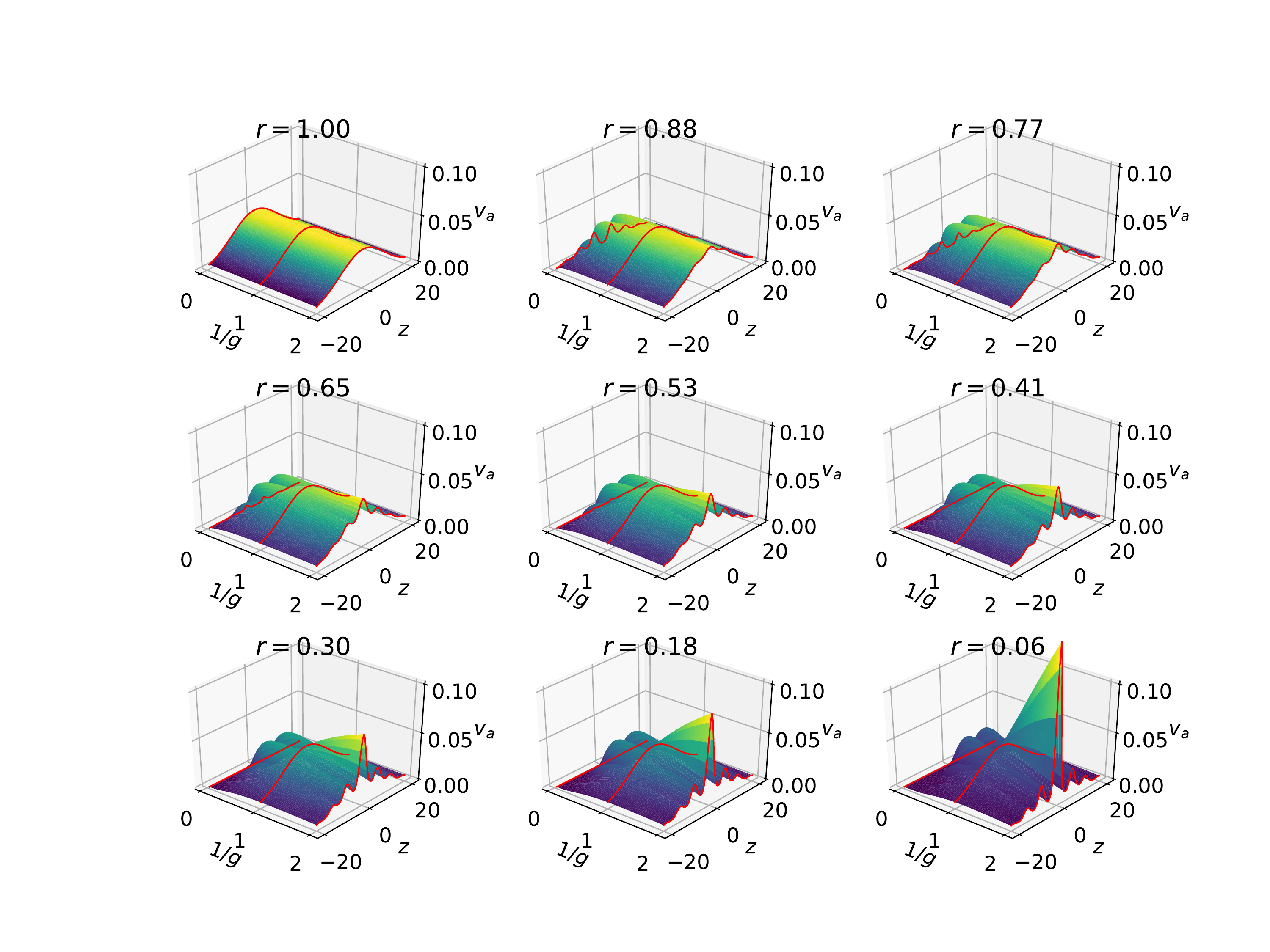}
		\caption{Three-dimensional evolution of the non-local part of a relative Gaussian backscattering interaction $v_a(z)$ with typical width $a=10$ (all length scales in units of $\Lambda^{-1}$) as a function of the relative coordinate $z$ and the inverse LL parameter $1/g$. The quantity $r$ on each graph represents the ratio of the running cutoff to the initial cutoff. $1/g<1$ corresponds to attractive interactions, while $1/g>1$ corresponds to repulsive interactions in the LL. The central red line at $g=1$ is here to illustrate that no renormalization occurs for the non-interacting case $g=1$, as expected. 
}
		\label{graph_renormalisation_simple}
\end{figure*}

In order to make progress, we make the further simplifying assumption that $\bar{V}(x_G,z)$ is the product of a center of mass contribution $V_L(x_G)$ (localized around the position $x_G=0$),  and a relative contribution $v_a(z)$ (labeled below as the relative XBH) which specifies the hopping range. According to the previous section, only $v_a(z)$ is modified under renormalization.
We first illustrate our result with a (relative) XBH amplitude $v_a(z)$ which contains a single central peak at $z=0$. Although any form for $v_a(z)$ with a characteristic length scale $a$ can be employed, for specificity, we choose a Gaussian form:
\begin{eqnarray}\label{Gaussian_form_potential}
v_a(z)=\frac{1}{\sqrt{2\pi a^2}}\exp\left(-\frac{z^2}{2a^2} \right).
\end{eqnarray}
The range $a$ is chosen to be one order of magnitude larger than the inverse momentum cutoff ($a=10 \Lambda^{-1}$). In Fig.~\ref{graph_renormalisation_simple}, we display the evolution of the shape of the relative interaction $v_a(z)$ at different steps of the RG procedure. There, the ratio $r$ displayed on top of each graph represents the ratio between the running momentum cutoff and the initial momentum cutoff (or alternatively, the ratio between the running energy cutoff and the initial energy cutoff since the dispersion in the LL is assumed to be linear). One notices that very early in the renormalization procedure ($r=0.88$), $v_a(z)$ already undergoes substantial modifications: oscillations appear for values of the coupling parameter $g$ which correspond to attractive interactions in the LL, and the overall interaction $v_a(z)$ is tilted upwards toward the region $g<1$ (repulsive interactions). These oscillations correspond to the spacing between minima/maxima of the exponent $\alpha(g,\Lambda,z)$, itself governed by the oscillations of the Bessel function $J_0$, as discussed when describing Fig.~\ref{graph_renormalization}. For $r$ ranging between $0.65$ and $0.41$, such oscillations propagate over the whole range of attractive and repulsive values of $g$. For $g>1$, one also notices the formation of two ``bumps'', which stay stable upon further renormalization. In the opposite case of repulsive interactions, $g < 1$, a small peak rises at $z=0$ and grows considerably, getting higher and narrower as one moves forward in the renormalization treatment. This is well illustrated in the lower right panel of  Fig.~\ref{graph_renormalisation_simple}, which corresponds to $r=0.06$. One readily sees that the height of this peak is larger when repulsive interactions are stronger ($1/g\sim 2$).
 
\begin{figure}[tbp]
	\centering
		\includegraphics[width=0.7\textwidth]{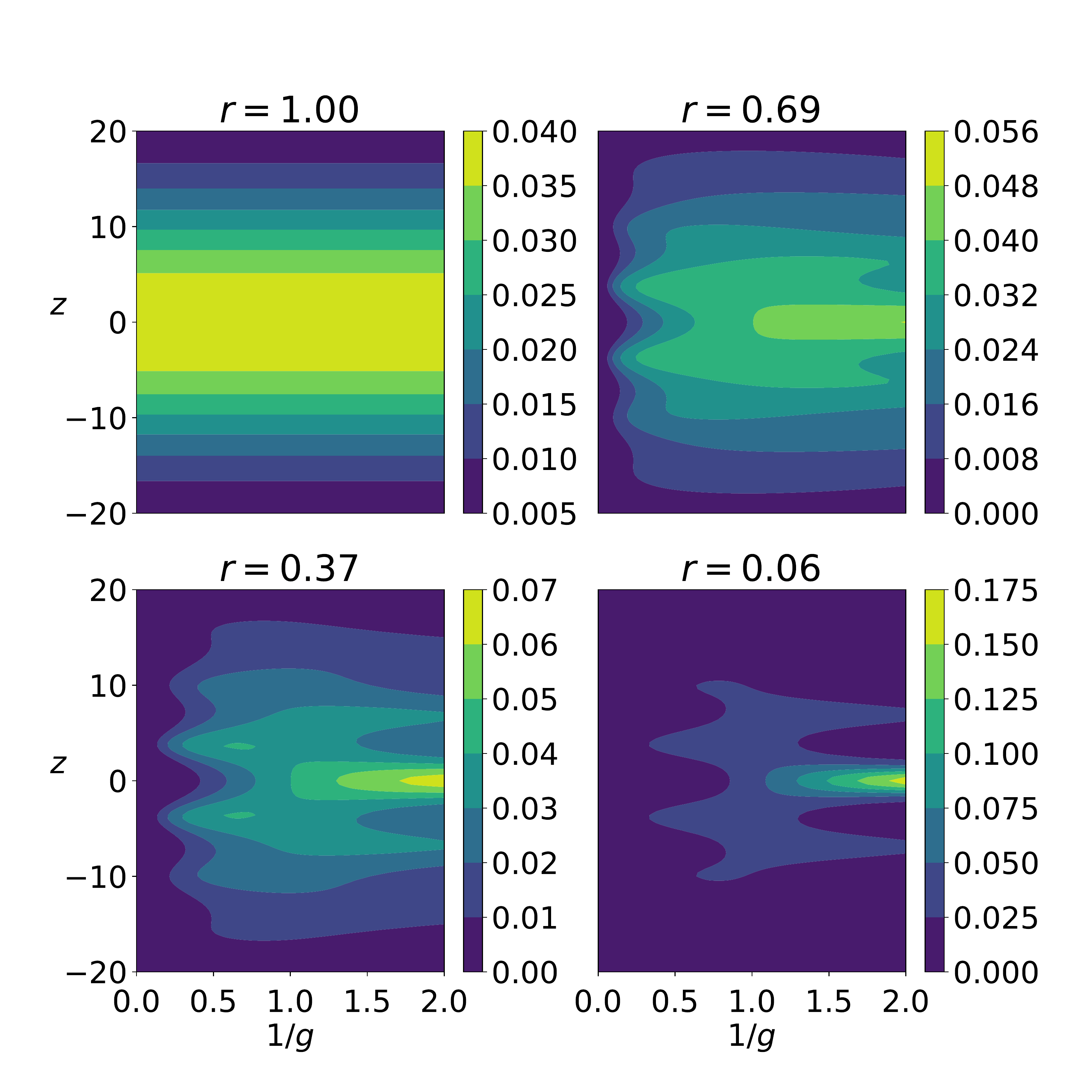}
		\caption{Two-dimensional plot of the relative Gaussian interaction $v_a(z)$, as a function of the relative coordinate $z$ and the inverse LL parameter $1/g$, for $r= 1.0,~ 0.69, ~0.37, ~0.06$.}
		\label{two_d_simple}
\end{figure}

As the renormalization procedure is taken  further yet (not shown), say $r=0.001$, the scale of the plots has to be changed in order to visualize the totality of the central peak at $z=0$ for $g<1$, while the structures/oscillations for attractive values of the interaction ($g>1$) can no longer be noticed. This indicates that if the renormalization procedure is pursued until an energy cutoff is reached, corresponding to ultra low temperatures, an extended interaction $v_a(z)$ converges towards zero for $g>1$, while it becomes a delta function scalar potential for repulsive interactions. While this result is somewhat expected, to our knowledge no indication or proof of such statement was presented so far in a quantitative manner in the literature.   

The second general comment that one can make is that, even when the renormalization is pushed to moderate values of the ratio of the running cutoff to its initial value (say  $r\sim 0.1$), the interaction landscape is already drastically modified, severely departing from its initial Gaussian shape of Eq.~(\ref{Gaussian_form_potential}). Interacting electrons thus experience a relative interaction $v_a(z)$ which is strongly altered, bearing new maxima and minima as a function of the relative distance $z$.  As a guide to the eye, we have indicated in Fig.~\ref{graph_renormalisation_simple} with red lines the amplitude of $v_a(z)$ for three values of the LL parameter: $g=+\infty$, $g=1$, and $g=0.5$. This confirms in particular that for a Fermi liquid ($g=1$), no renormalization occurs. Conversely, $g=+\infty$ yields a straight line signaling the full suppression of $v_a(z)$, while for repulsive interactions ($g=0.5$) the oscillations associated with $\alpha(g,\Lambda,z)$ are clearly visible. 

As a final comment, let us emphasize that we could check that these results are qualitatively robust when using smooth cutoff functions, in the spirit of \cite{knops_80}. Details of the calculations and results are provided in  \ref{softcutoff_appendix}. This procedure does not significantly affect our results, the main difference appearing in the case of a soft cutoff function being the washing out of the oscillations observed in Fig.~\ref{graph_renormalisation_simple}.

As the interpretation of the 3D plots may be confusing, and in order to simplify the comparison with other instances of extended potential, we also provide in Fig.~\ref{two_d_simple} a two-dimensional version of the results of Fig.~\ref{graph_renormalisation_simple}. At $r=0.69$, one still identifies the general Gaussian shape of the relative interaction, except for the fact that $v_a(z)$ has almost completely vanished for $1/g< 0.2$ (attractive LL parameter), while a clear maximum develops for $1/g>1$ (repulsive LL parameter). Continuing the renormalization procedure to $r=0.37$, one notices that the amplitude of  $v_a(z)$ is further modified, as features away from $z=0$ have somewhat faded, while a sharp central peak at $z=0$ emerges for $1/g>1$, being the only surviving feature when ultimately reaching $r=0.06$.  Furthermore, the presence of oscillations in $z$ for a given value of the LL parameter clearly appears in Fig.~\ref{two_d_simple}, where one recognizes, e.g. a double barrier structure for $1/g\sim 0.5$ (attractive interactions) or a triple peak structure for $1/g>1.3$ (repulsive interactions). It is important to stress out that this renormalized interaction $v_a(z)$ should in principle be used to compute transport properties such as the electric current and noise characteristics, which may have significant consequences if the experimental constraints require to stop the renormalization procedure at intermediate values of $r$. In the discussion of Sec. \ref{discussion}, we examine physical systems and experimental conditions which justify stopping the renormalization procedure around $r\sim 0.06$, thus highlighting the relevance of the present study.

\subsection{Application to an extended double barrier backscattering interaction} \label{extended_double_barrier}

\begin{figure}[tbp]
	\centering
		\includegraphics[width=0.7\textwidth]{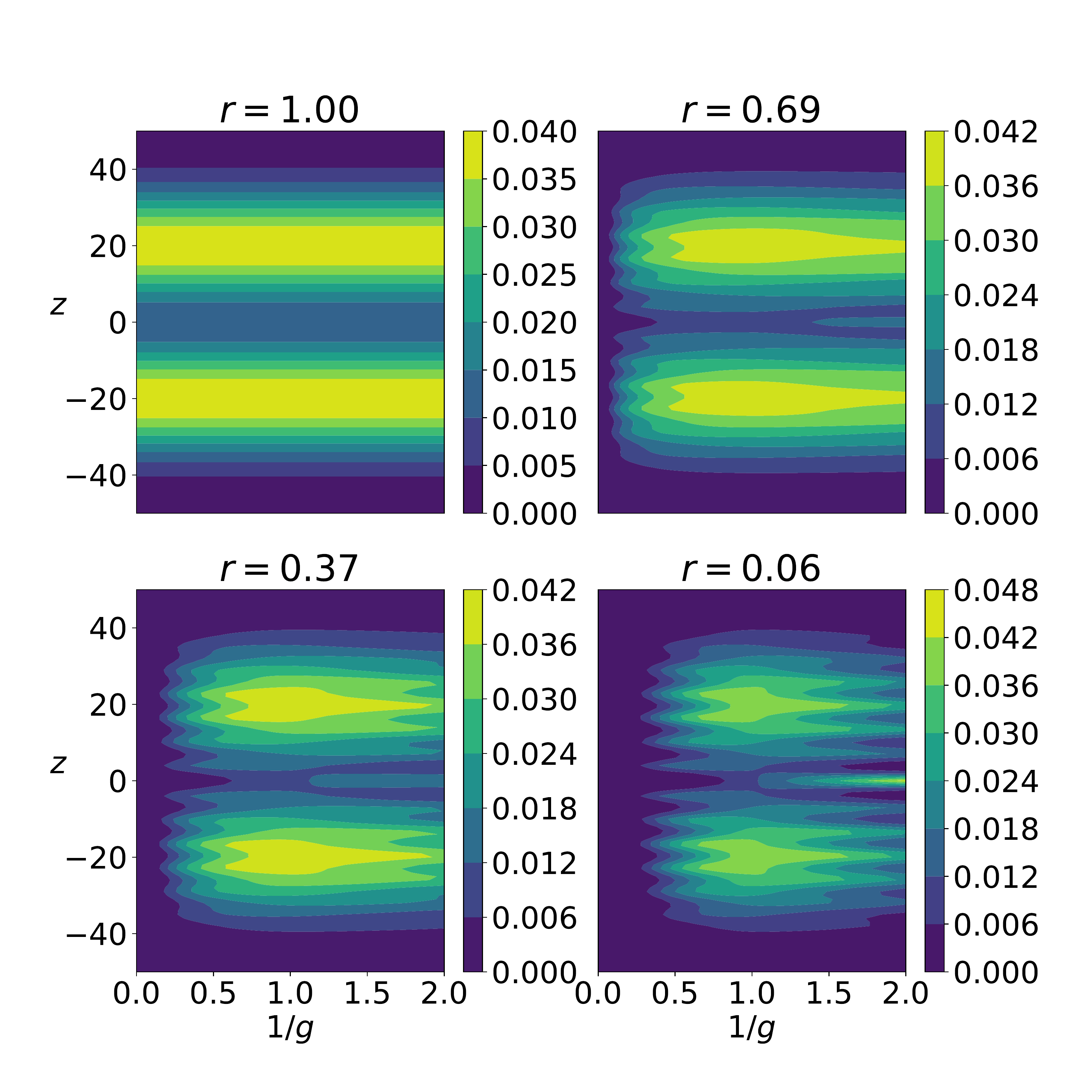}
		\caption{Two-dimensional plot of the evolution of the amplitude of the non-local part of a relative hopping interaction $v_a(z)$ which consists of two well separated Gaussians, Eq.~(\ref{2_gaussian}). Units and parameters are chosen to be the same as in Fig.~\ref{two_d_simple}.}
		\label{graph_renormalisation_double}
\end{figure}

For completeness, we also consider the evolution under renormalization of a relative hopping interaction $v_a(z)$ which has maxima away from $z=0$. The corresponding results are displayed in Fig.~\ref{graph_renormalisation_double}. We choose for specificity a profile of $v_a(z)$ which consists of a superposition of two Gaussians, with maxima shifted at $\pm 2a$ (keeping $a=10 \Lambda^{-1}$): 
\begin{eqnarray}
v_a(z)&= \bigl(2\pi a^2\bigr)^{-1/2} \Bigl[ \exp\bigl(-(z+2a)^2/2a^2\bigr) + \exp\bigl(-(z-2a)^2/2a^2\bigr)\Bigr].
\label{2_gaussian}
\end{eqnarray}
Such a hopping interaction could in principle represent  the XBH components of the double saddle point Hamiltonian of a double quantum point contact. 

We observe that upon renormalization, the evolution of the double barrier relative interaction is quite slow compared to the one of the single Gaussian of Fig.~\ref{two_d_simple}. This is because the amplitude $v_a(z)$ is small but non-zero close to $z=0$ due to the double barrier structure. However, for rather small values of $1/g$ (attractive interaction) the relative XBH amplitude is suppressed, as expected from the structure at the same parameters in Fig.~\ref{graph_renormalization}. One has to push the renormalization to a ratio $r=0.06$ (about $1/16$th of the initial momentum cutoff) in order to distinguish the rise of a delta-like potential at $z=0$ and $g<1$. At this step of the renormalization, the initial double barrier potential $v_a(z)$ has evolved into a triple barrier structure, with a central and dominant peak at $z=0$.

\begin{figure}[tbp]
	\centering
		\includegraphics[width=0.7\textwidth]{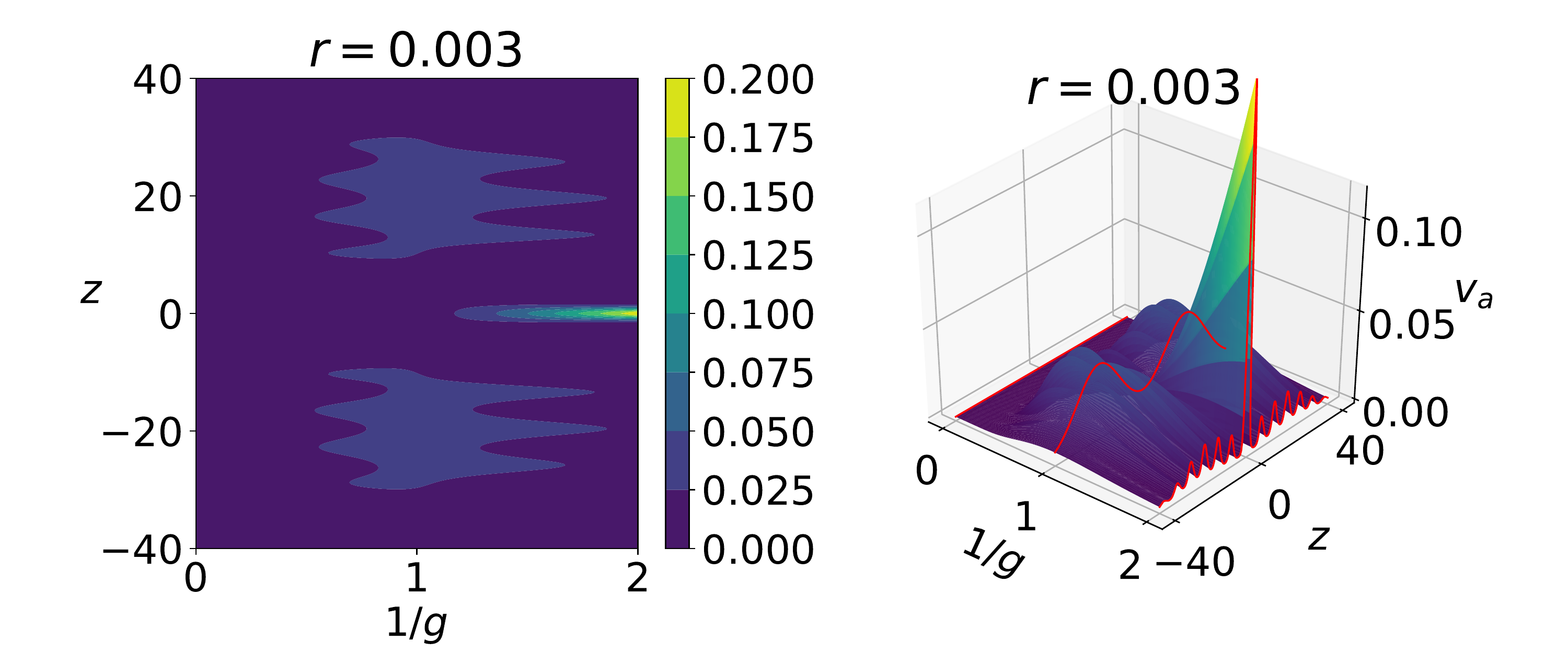}
		\caption{Two-dimensional (left) and three-dimensional (right) plots of the amplitude of the relative double Gaussian XBH amplitude $v_a(z)$ [see Eq.~(\ref{2_gaussian})] after substantial renormalization has been operated, $r=0.003$. Units and parameters are chosen to be the same as in Fig.~\ref{two_d_simple}. The red line at $g=1$ indicates that no renormalization occurs for the Fermi liquid case.}
		\label{graph_renormalisation_double_end}
\end{figure}

For illustrative purposes, we display in Fig.~\ref{graph_renormalisation_double_end} the evolution of $v_a(z)$ once $r=0.003$ has been reached. The profile is, as expected, dominated by a delta-like peak at $z=0$ with much suppressed side peaks at $z\neq 0$. For $1/g<1$, one still recognizes the reminiscence of a double barrier XBH, albeit with reduced amplitude. 

%%%%%%%%%%%%%%%%%%%%%%%%%%%%%%%%%%%%%%%%%%%%%%%%%%%%%%%%%%%%%%%%%%%%%%%%

\section{Perturbative calculation of the current} \label{perturbative_current}

Previous works on extended impurities in the fractional quantum Hall effect, modeled as a chiral Luttinger liquid \cite{chevallier_10b}, have shown that they typically lead to significant changes in the transport properties. The same kind of behavior is expected here in the case of non-chiral LL. Indeed, we showed in the previous sections that in the limit of a fully  converged flow the extended backscattering Hamiltonian reduces to a delta-like scalar potential, therefore leading back to the standard power-law dependence \cite{kane_92c}. However, at an intermediate step of the RG flow, we expect to see a non-universal behavior departing from these power-law results, and depending on the details of the extended backscattering potential. We thus propose some steps for the perturbative calculation of the current in the presence of an arbitrary XBH. We will stay general in this section and restrict ourselves to  the principle of the calculation.

The total current corresponds to the maximal current $I_{max}\equiv e^2gV/h$ of a pure LL, from which one subtracts the backscattering current $I_B$ which we compute here. In order to include the bias potential difference between the two extremities of the wire, the impurity action is modified in order to include a ``vector potential" $A(\tau)$ (the real time voltage being given by $V(t) = \partial_t A(t)$):
\begin{eqnarray}
S_{\rm{imp}}&=\int d\tau\,dx_G\,dz\,\eta^{-1}\bar{V}(x_G,z) \nonumber \\
&\quad\times \cos\left\{ \sqrt\pi\bigl[ \theta\left(x_G+ z/2,\tau\right)+\theta(x_G- z/2,\tau) +\phi\left( x_G+ z/2,\tau\right) \right. \nonumber \\
&~~ \quad\quad\quad \left. - \phi\left( x_G- z/2,\tau\right)\bigr]  +2k_F x_G+g A(\tau)\right\},
\end{eqnarray}
where the prefactor $g$ in front of $A(\tau)$ originates from the fact that the backscattering occurs in a correlated state of matter.~\cite{kane_92b,kane_92c} 

We follow closely the approach described in the Appendix of~\cite{kane_92c}, which expands the imaginary time partition function in terms of powers of the backscattering Hamiltonian. We define the zeroth ($Z_0$), first ($Z_1$) and second order ($Z_2$) terms of this expansion by
\begin{eqnarray}
Z&=\left\langle e^{-(S_0+S_{\rm{imp})}}\right\rangle \nonumber \\
&= \left\langle e^{-S_0}\right\rangle + \left\langle -S_{\rm{imp}}\right\rangle_0+\frac{1}{2!}\left\langle S_{\rm{imp}}S_{\rm{imp}}\right\rangle_0+\ldots \nonumber \\
&= Z_0+Z_1+Z_2+\ldots
\end{eqnarray}
The calculation of $Z_1$ follows closely the calculation performed in the perturbative RG treatment, except that instead of integrating over the fast degrees of freedom only, one integrates over all fields:
\begin{eqnarray}\label{Z1}
Z_1=-\int d\tau\,dx_G\,dz\, & \eta^{-1} \bar{V}(x_G,z)  \cos\bigl[2k_Fx_G+g A(\tau)\bigr] \nonumber \\ 
&\times e^{ -\pi\bigl[ G^{\theta\theta}(0,0)+G^{\phi\phi}(0,0)+G^{\theta\theta}(z,0)-G^{\phi\phi}(z,0)\big]}.
\end{eqnarray}
In \ref{perturbative_appendix} we show that the argument of the exponential is $-\infty$, so that $Z_1=0$. This result is indeed consistent with the perturbative result of~\cite{kane_92c} for the delta function impurity potential. 

Next, we focus on $Z_2$, which we write in the following form: 
\begin{eqnarray}\label{Z2}
Z_2&=\frac 18\int d\tau\, dx_G\, dz\, d\tau'\, dx'_G\,dz'\, \eta^{-2} \bar{V}(x_G,z) \bar{V}(x'_G,z') \nonumber \\
&\quad\sum_{\sigma,\sigma'=\pm} \biggl\langle  e^{
	 i \sigma \sqrt\pi\bigl[ \theta\left(x_G+ z/2,\tau\right)+\theta\left( x_G- z/2,\tau \right)    + \phi\left( x_G+ z/2,\tau\right)- \phi\left( x_G- z/2,\tau\right) \bigr]} \nonumber \\
&\qquad\times e^{ i \sigma' \sqrt\pi\bigl[ \theta\left(x'_G+z'/2,\tau'\right)+\theta\left(x_G- {z'}/2,\tau'\right)+ \phi\left( x'_G+ {z'}/2,\tau'\right)- \phi\left( x'_G- {z'}/2,\tau'\right)\bigr]} \nonumber \\
&\qquad\times  e^{ i \sigma \Bigl(2k_F x_G+g A(\tau)\Bigr)} e^{ i \sigma' \Bigl(2k_F x'_G+g A(\tau') \Big)} \biggr\rangle .
\end{eqnarray} 

In \ref{perturbative_appendix}, we show that only $\sigma=-\sigma'$ gives a non-zero contribution. The Gaussian integrals over exponentiated fields are computed in a similar manner as for the integration of the fast degrees of freedom and the details of this calculation are provided in the Appendix. We thus obtain the final result for $Z_2$:
\begin{eqnarray}
Z_2 &= 
\frac{1}{4}
\int d\tau\, dx_G\, dz\,d\tau'\,dx'_G\,dz'\, \eta^{-2}\bar{V}(x_G,z)   \bar{V}(x'_G,z') \nonumber \\
& \times  \mathcal J_-(x_G,x'_G,z,z',\tau-\tau')  \cos\bigl[2k_F(x_G-x'_G)+g A(\tau)-g A(\tau')\bigr],
\end{eqnarray}
where $\mathcal J_-(x_G,x'_G,z,z',\tau-\tau')$ is computed in \ref{perturbative_appendix}.

The (imaginary time) backscattering current is then obtained by taking derivatives of $Z_2$ with respect to the vector potential $A(\tau)$:
\begin{eqnarray}
I_B(\tau)&= 
\frac{g}{2} \int d\tau'\, dx_G\, dz\, dx'_G \,dz'\, \eta^{-2}\bar{V}(x_G,z)  \bar{V}(x'_G,z')  \nonumber \\
& \times \mathcal J_-(x_G,x'_G,z,z',\tau-\tau') \sin\bigl[2k_F(x_G-x'_G)+g A(\tau)-g A(\tau')\bigr].
\label{imaginary_current}
\end{eqnarray}
In the case of a localized impurity where $\bar{V}(x_G,z)=W\delta(x_G)\delta(z)$ with constant amplitude $W$, the backscattering current in real time reduces to
\begin{eqnarray}
I_B(t)&=
\frac{g}{2} W^2\int_{-\infty}^t dt'\, \eta^{-2}\sin\bigl[g A(t)-g A(t')\bigr] \frac{P^>(t-t')-P^<(t-t')}{i},
\end{eqnarray}
where $P^{>(<)}(t-t')$ is the analytic continuation of $\mathcal J_-(0,0,0,0, \tau-\tau')$ for $\tau=+(-)it$, in accordance with~\cite{kane_92c}.

For a general form of the impurity potential, the real time current can be obtained by changing the contour of the imaginary time integral in Eq.~(\ref{imaginary_current}) to the new contour for $t'=-i\tau'$ running from $-\infty$ to $t$, then back to $-\infty+i\beta$. We thus obtain a general expression for the current in real time
\begin{eqnarray}
I_B(t)&=
\frac{g}{2}
 \int_{-\infty}^t dt'\,dx_G\, dz\, dx'_G\, dz'\, \eta^{-2} \bar{V}(x_G,z)  \bar{V}(x'_G,z')  \nonumber \\
&\times \sin\bigl[2k_F(x_G-x'_G)+g A(t)-g A(t')\bigr] \nonumber \\
&\times \frac{ P^>(x_G,x'_G,z,z',t-t')-P^<(x_G,x'_G,z,z',t-t')}{i},
\label{final_perturbative}
\end{eqnarray}
where $P^{>(<)}(x_G,x'_G,z,z',t-t')$ is the analytic continuation of $\mathcal J_-(x_G,x'_G,z,z', \tau-\tau')$ for $\tau=+(-)it$.

In the event that the renormalization procedure has to be stopped because a cutoff such as the lowest Matsubara frequency has been reached, one should in principle insert the renormalized relative XBH obtained in Sec.~\ref{relative_potential_section}, into the perturbative calculation of the current [see Eq.~(\ref{final_perturbative})] in order to make experimental contact. Such a calculation would also involve an analytic continuation procedure, likely to require advanced numerical techniques such as maximum entropy methods, which go beyond the scope of this paper.

%%%%%%%%%%%%%%%%%%%%%%%%%%%%%%%%%%%%%%%%%%%%%%%%%%%%%%%%%%%%%%%%%%%%%%%%
%%%%%%%%%%%%%%%%%%%%%%%%%%%%%%%%%%%%%%%%%%%%%%%%%%%%%%%%%%%%%%%%%%%%%%%%
%%%%%%%%%%%%%%%%%%%%%%%%%%%%%%%%%%%%%%%%%%%%%%%%%%%%%%%%%%%%%%%%%%%%%%%%
%%%%%%%%%%%%%%%%%%%%%%%%%%%%%%%%%%%%%%%%%%%%%%%%%%%%%%%%%%%%%%%%%%%%%%%%

\section{Discussion} \label{discussion}

The main motivation of this study was to inquire whether the shape of the XBH amplitude at intermediate steps of the renormalization procedure should be kept as a possible input for the calculation of the current when physical or experimental reasons command to stop the renormalization procedure. 

As indicated in most textbooks dealing with this method, a possible lower cutoff is the lowest Matsubara frequency $2\pi T_0$, where $T_0$ is the temperature at which the transport experiment is carried out. As we wish to stay in the quantum coherent regime, we choose an optimistic upper bound  $T_0=1 K$. For the one-dimensional system under study, we can consider either atomically defined wires such as metallic carbon nanotubes (for which the Fermi wave vector is proportional to the inverse of the lattice spacing), or artificially designed nanowires consisting of semiconductors (grown, etched, or 2D electron gases defined by neighboring metallic gates). 

\begin{figure}[tbp]
	\centering
		\includegraphics[width=0.7\textwidth]{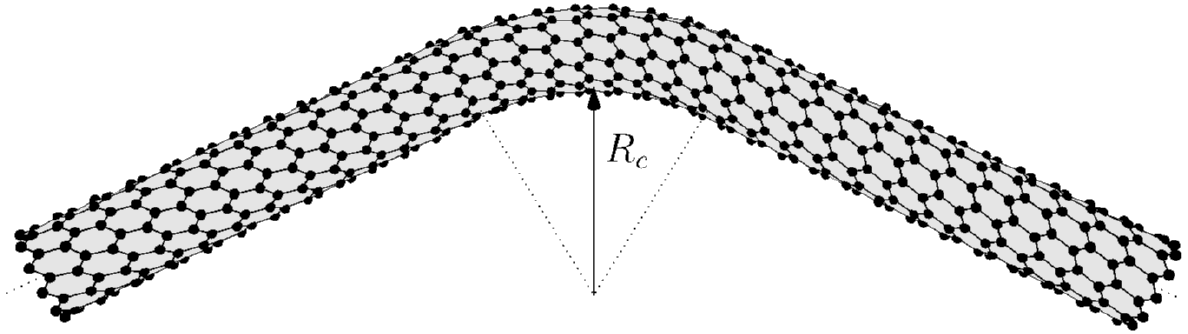}
		\caption{Schematic representation of a section of a carbon nanotube bent at a specific location, with a radius of curvature $R_c$. The bent region corresponds typically to the range of the XBH of our study.}
		\label{nanotube}
\end{figure}

Nanotubes have a rather large Fermi energy (as it is related to the lattice constant), which means that the ratio $r$ between the lower cutoff $T_0$ and the initial cutoff could be as small as $10^{-2}-10^{-3}$. Unless the renormalization procedure has to be stopped imperatively for physical reasons, or unless the extent of the relative XBH is quite large ($a\sim 100\Lambda^{-1}$), it is therefore less likely that the evolution of the relative XBH yields something much different than a delta function peak as these two energy scales are far apart. However, for metallic carbon nanotubes, an XBH can be generated when such nanotubes have a ``bend'' in a specific location, as illustrated in Fig.~\ref{nanotube}.~\cite{farajian_03} When the radius of curvature of the nanotube is of the order of the lattice constant, this sharp bend represents a localized impurity scalar potential, where the results of~\cite{kane_92a,kane_92b,kane_92c} apply. However, if the radius of curvature $R_c$ of this bend is several orders of magnitude larger than the lattice constant, one expects that the conditions for realizing an XBH are satisfied, pointing out to the relevance of this study.

On the other hand, semiconductor wires have a much smaller Fermi energy (corresponding to a temperature of a few hundred Kelvins) which can in principle be controlled by a back gate, allowing to lower the electron density. For these semiconductor nanowire systems, the momentum cutoff of the LL theory should correspond to the energy interval where the linear dispersion is valid. We thus choose for such systems an initial cutoff of the order of $100 K$. This means that the ratio between the final and initial energy cutoff could be somewhat low, but not so low as to blur the structures of the renormalized XBH. In our plots describing the evolution of the relative XBH amplitude, we concluded that even if a ratio $r= 0.06$ is reached -- a ratio which is consistent with the above estimate of the initial and final cutoff choices -- the relative XBH amplitude still contains non-trivial spatial structures when the renormalization procedure has to be stopped. Moreover, as semiconductor wires (which may have several conduction channels) are expected to have a LL parameter $g>0.5$ (as opposed to a LL parameter $g\sim 0.3$ for carbon nanotubes) we anticipate that when realistic lower momentum/frequency cutoffs are reached, the complex structures found for the renormalized XBH $\bar{V}(x_G,z)$ should constitute a realistic input for the computation of transport properties.      

For simplicity, we chose to present our results in the case of a spinless Luttinger liquid. However, one may argue that realistic experimental devices are more likely to be spinful. We want to stress out here that our derivation can be naturally extended to treat spinful Luttinger liquids. In this situation, one needs to introduce a separate interaction parameter for the spin and charge sector: $g_s$ and $g_c$ (note that following \cite{kane_92c}, non-interacting electrons now correspond to $g_c = g_s = 2$). Our derivations can be reproduced, leading to similar expressions upon replacing $g \to \frac{g_c + g_s}{4}$ and $g^{-1} \to g_c^{-1} + g_s^{-1}$. In particular, in the (most common) SU(2) symmetric case where one has $g_s=2$, our central result of Eq.~(\ref{eq:alpha}) becomes
\begin{eqnarray}
 \alpha_{\textrm{spinful}} (g,\Lambda,z)= \frac{1}{2}-g_c^{-1}+\frac{g_c^{-1}-\frac{g_c}{4}}{2}\bigl[1+J_0(\Lambda |z|)\bigr]\label{alpha},
\end{eqnarray}
This, however, has no significant effect on our results which stay qualitatively robust.

The last part of this work was devoted to the perturbative calculation of the current, which has the benefit of using the same formalism as the renormalization procedure. We were able to isolate the contributions to the imaginary time current, and to show that in the limit of a short ranged potential the result of~\cite{kane_92c} is recovered. For the case of an extended potential, the ultimate step for obtaining the current, and possibly the power law behavior of the conductance, is to perform an analytic continuation of the impurity correlator Kernel of Eq.~(\ref{j-final}). This allows in principle to compute the transport properties - such as the real time current, the conductance, and possibly the noise - corresponding to the relative XBH that is obtained when the renormalization procedure has to be stopped, invoking experimental conditions.  

Note that there are some restrictions about the range of  $\bar{V}(x_G,z)$, which by construction describes an instantaneous transfer of right/left moving  fermions into left/right moving fermions. The distance $z$ which separates the electron destruction and creation location can of course be larger than the lattice constant, but physically  should be smaller than the product of the Fermi velocity with the tunneling time (associated with backscattering) of such electron transfer processes~\cite{landauer_94}.

%%%%%%%%%%%%%%%%%%%%%%%%%%%%%%%%%%%%%%%%%%%%%%%%%%%%%%%%%%%%%%%%%%%

\section{Conclusion} \label{conclusion}

In this work, we have revisited a ``classic'' problem of transport in one-dimensional strongly correlated systems, albeit in the presence of an extended backscattering Hamiltonian. To our knowledge, studies have so far mostly focused on short-ranged, delta-like impurity potentials. In this different situation, say, a right moving electron at location $x$ can be converted into a left moving electron at a nearby location $x'$ at a distance corresponding to several inverse Fermi momenta. We focused on kinetic hopping interactions which can be cast in full generality into functions of a center of mass coordinate $x_G=(x+x')/2$ (XBH center of mass with extent $L$) and of a relative coordinate $z=x-x'$ (XBH relative coordinate with extent $a$). We used a momentum shell RG treatment of the impurity action which shows that while $\bar{V}(x_G,z)$  is unchanged with respect to $x_G$, the relative $z$-dependence of this XBH is modified under renormalization.
Fundamental changes in shape are described by a renormalization flow equation involving the relative coordinate $z$ with a strong dependence on the repulsive/attractive nature of the LL. This constitutes the core result of this study. 

We illustrated this result by monitoring the evolution of the interaction with respect to this relative coordinate, starting from a relative XBH with a single maximum at $z=0$ which is typically much wider than the inverse momentum cutoff (which would also correspond to a length scale of the order of the lattice constant). We confirmed the conjecture stating that if the renormalization procedure is pursued until zero temperature (or a very small momentum cutoff) is reached, the initial relative XBH amplitude becomes ultimately a delta-like function peak centered at $z=0$. This turns out to be true even when the relative XBH has no initial maximum at this location, as illustrated with a double barrier relative XBH amplitude which contains two well separated maxima symmetrically placed around $z=0$.  

There are many interesting extensions of our present work. First, a natural direction to explore would be to incorporate higher order contributions to the RG flow. Indeed, we expect that, unlike local impurity potentials which do not lead to a renormalization of bulk quantities, the presence of a spatially extended potential might lead to a renormalization of the Luttinger liquid parameter, involving contributions at higher order, beyond what we considered here.

Second, it would be interesting to extend the present formalism to chiral Luttinger liquids, used to describe the edge states of the fractional quantum Hall effect.~\cite{laughlin_99,stormer_99a,stormer_99b} There, for simple Laughlin fractions, the chiral excitations which propagate on opposite edges of the quantum Hall bar give rise to the tunneling of Laughlin quasiparticles - rather than that of electrons - from one edge to the other, at the location of the quantum point contact. As mentioned in the introduction,  the calculation of the current and noise can be performed in the presence of an XBH in the Poissonian limit.~\cite{chevallier_10b} However, it would also be informative here to perform a renormalization group treatment of the XBH in this context, which amounts to studying the scaling dimension of the tunneling operator, the latter being expressed as exponentials of the difference between the two chiral LL bosonic fields. 

Finally,  in~\cite{kane_92a,kane_92b,kane_92c} a duality correspondence between weak backscattering and strong backscattering (where the LL wire is effectively split into two semi-infinite LL wires, with a single tunneling location between the two extremities) was identified. The simultaneous exchange of the bosonic fields $\theta$ and $\phi$, with the replacement $g$ by $1/g$ allows to extract directly all information for the strong backscattering case from the weak backscattering results. One could possibly exploit this duality once again to treat electron tunneling between two semi-infinite LLs  over an extended  tunneling region (where fermion creation/destruction locations are distributed in the vicinity of the two contacts), in an analogous manner as the present study for weak backscattering, using the same duality transformation as in Refs.~\cite{kane_92a,kane_92b,kane_92c}.

\ack
The project leading to this publication has received funding from Excellence Initiative of Aix-Marseille University - A*MIDEX, a French "Investissements d'Avenir" program. A.V.L. acknowledges the support from the RFBR Grant No. 18-02-00642A and from the Government of the Russian Federation (Agreement 05.Y09.21.0018).

%%%%%%%%%%%%%%%%%%%%%%%%%%%%%%%%%%%%%%%%%%%%%%%%%%%%%%%%%%%%%%%%%%%%%

\appendix

\section{Detailed renormalization procedure} \label{detailed_RG}

The first step of the renormalization procedure consists in decomposing the fields  $\phi$ and $\theta$ in their fast and slow components: $\phi=\phi'+\tilde\phi$ and $\theta=\theta'+\tilde\theta$, in such a way that the fast components $\tilde\phi$ and $\tilde\theta$ are non-zero for $\Lambda/b\leq |\vec p|\leq \Lambda$, and zero otherwise. The  free action then becomes
\begin{eqnarray}
S_0(\phi,\theta)=S_0(\phi',\theta')+S_0(\tilde\phi,\tilde\theta)~.
\end{eqnarray}
The technical aspect of averaging  the impurity action over fast degrees of freedom reduces to the calculation of Gaussian integrals over the fields $\theta$ and $\phi$. This is explained in \ref{average_exponentials}. We obtain the following results:
\begin{eqnarray}
&&\Bigl\langle \exp \pm \Bigl( i\sqrt\pi\Bigl[ \tilde{\theta}\left(x_G+ z/2,\tau\right)+\tilde{\theta}\left(x_G- z/2,\tau\right) \Bigr.\Bigr.\Bigr.\nonumber\\
&&~~~~~~~\Bigl.\Bigl.\Bigl.+\phi\left( x_G+ z/2,\tau\right)- \phi\left( x_G- z/2,\tau\right)\Bigr]\Bigr)\Bigr\rangle\nonumber\\
&&=\exp\Bigl( -\pi\Bigl[ \tilde{G}^{\theta\theta}(0,0)+\tilde{G}^{\phi\phi}(0,0)+\tilde{G}^{\theta\theta}(z,0)-\tilde{G}^{\phi\phi}(z,0)\Bigr]\Bigr)~,
\end{eqnarray}
where we introduced the fast modes Green's functions as follows: 
\begin{eqnarray}
      \tilde G^{\phi\phi}(z,0)=\int\limits_{\Lambda/b\leq|\vec p|\leq \Lambda}
      \frac{d^2 p}{(2\pi)^2}\;G^{\phi\phi}(\vec p)\,e^{iqz},
\end{eqnarray}
and similarly for $\tilde G^{\theta\theta}$. 
This allows to write the backscattering action after integration of the fast degrees of freedom:
\begin{eqnarray}
      \bigl\langle S_\mathrm{imp}\bigr\rangle_f&= \int d\tau\, dx_G\, dz\, \eta^{-1}V_L(x_G)
      v_a(z) \nonumber \\
 &\times\exp\Bigl(-\pi\Bigl[\tilde G^{\theta\theta}(0,0)
      +\tilde G^{\phi\phi}(0,0)+\tilde G^{\theta\theta}(z,0)-\tilde G^{\phi\phi}(z,0)\Bigr]\Bigr)  \nonumber \\
      &\times\cos\Big(
      \sqrt\pi\Bigl[\theta^\prime(x_G+ z/2,\tau)+\theta^\prime(x_G- z/2,\tau)\Bigr.  \nonumber \\
      &\Bigl.\qquad\qquad+\phi^\prime(x_G+ z/2,\tau)-\phi^\prime(x_G- z/2,\tau)\Bigr]+2k_Fx_G\Bigr).
      \label{simpmoy}
\end{eqnarray}

The last step of the renormalization procedure consists in a rescaling step allowing to recover the initial cutoff. The free action  $S_0$ is taken as a reference and one wishes that it remains invariant, thus specifying the rescaling of parameters. 

The integrals in Eq.~(\ref{simpmoy}) are defined for $|\tau|>b\Lambda_\tau$, $|x|>b\Lambda_x$ and $|z|>b\Lambda_z$ where the cutoffs $\Lambda_\tau,\Lambda_x,\Lambda_z$ correspond to the variables
$\tau,x,z$.

In order to reestablish the initial cutoffs, we proceed to a rescaling of the variables:
\begin{eqnarray}
\begin{array}{clrr} \tau'=b^{-1}\tau,& x'=b^{-1}x,& z'=b^{-1}z\end{array},
\label{changement}
\end{eqnarray}
as well as the parameters $a$, $L$ and $\eta$. As the product $\vec{p}\cdot\vec{x}$ is dimensionless,  the momentum/frequency vector $\vec{p}=(q,\omega)$ becomes $\vec{p'}= b\vec{p}$, in particular $k_F$ is rescaled.
As we wish to conserve the algebraic structure of the cosine, the fields $\theta$ and $\phi$ are not rescaled themselves, and the rescaling occurs only through their arguments:
\begin{eqnarray}
\theta''(\tau',x'):= \theta'(\tau,x)=\theta'(b\tau',bx').
\end{eqnarray}
The impurity action then becomes:
\begin{eqnarray}
\bigl\langle S_{\mathrm{imp}} \bigr\rangle_f &= \int d\tau'\, dx'_G\, dz'\, b^{2}\eta^{-1}V_{bL'}(bx'_G)v_{ba'}(bz')   \nonumber \\
&\times\exp\Bigl( -\pi\Bigl[ \tilde G^{\theta''\theta''}(0,0)+\tilde G^{\phi''\phi''}(0,0)+\tilde G^{\theta''\theta''}(z',0)-\tilde G^{\phi''\phi''}(z',0)\Bigr] \Bigr) \nonumber \\
&\times\cos\Bigl( \sqrt{\pi}\Bigr[ \theta''(x'_G+{ z'}/{2},\tau')+\theta''(x'_G-{z'}/{2},\tau')\Bigr.\Bigr. \nonumber \\
&\qquad\quad\Bigl.+\phi''(x'_G+{z'}/{2},\tau')-\phi''(x'_G-{z'}/{2},\tau') + 2k_F'x'_G\Bigr).  
\label{S_impurity}
\end{eqnarray}
The fast modes Green's functions are also modified, for instance $\tilde G^{\phi\phi}$ 
becomes after rescaling:
\begin{eqnarray}
\tilde G^{\phi''\phi''}(z',\tau')&=\int\limits_{\Lambda\leq |{\vec p}^{\prime}|\leq b\Lambda}\frac{d^2p'}{(2\pi)^2}\frac{g^{-1}}{{\vec p}^{\prime 2}}e^{i\vec{p}'\cdot\vec {z}'}.
\end{eqnarray}
Focusing now solely on the relative interaction $v_a(z)$, its renormalized version $v_{a'}(z',b)$ reads:
\begin{eqnarray}
v_{a'}(z',b)&=&b  v_{ba'}(bz') \exp\Bigl( -\pi\Bigr[ \tilde G^{\theta''\theta''}(0,0)+\tilde G^{\phi''\phi''}(0,0)\Bigr.\Bigr.\nonumber\\
&&\qquad\qquad\qquad\Bigl.\Bigl.+\tilde G^{\theta''\theta''}(z',0)
-\tilde G^{\phi''\phi''}(z',0)\Bigr]\Bigr).
\label{vreno}
\end{eqnarray}
For a given $|z'|$ and $b=1+\epsilon$ (with $0<\epsilon\ll 1$), assuming a Gaussian form for $v_{ba'}(bz')$ as in Eq.~(\ref{Gaussian_form_potential}),
one then has
\begin{eqnarray}
bv_{ba'}(bz')=\frac{1}{\sqrt{2\pi a'^2}}\exp\biggl( -\frac{z'^2}{2a'^2}\biggr).
\end{eqnarray}
The fast modes Green's functions $\tilde G$ can be expressed as Bessel functions by performing the Fourier integration in polar coordinates.
They are integrated on the shell \rm{$\Lambda\leq |\vec{p}|\leq b\Lambda$.} The following integral representation of the zeroth order Bessel function
\begin{eqnarray}
J_0(z)=\frac 1\pi \int_0^\pi d\varphi\, \cos(z\cos \varphi),
\end{eqnarray}
leads to
\begin{eqnarray}
\tilde G^{\phi''\phi''}(z',0)&=\frac{g^{-1}}{2\pi}\int_{\Lambda}^{b\Lambda}dp\, \frac{J_0(p|z'|)}{p}\\
\tilde G^{\theta''\theta''}(z',0)&=\frac{g}{2\pi}\int_{\Lambda}^{b\Lambda}dp\, \frac{J_0(p|z'|)}{p}\\
\tilde G^{\phi''\phi''}(0,0)&=\frac{g^{-1}}{2\pi}\ln b\\
\tilde G^{\theta''\theta''}(0,0)&=\frac{g}{2\pi}\ln b\, ~.
\end{eqnarray}
This yields:
\begin{eqnarray}
v_{a'}(z',b)&=v_{a'}(z')b^{-1/g} \exp \biggl( \frac{g^{-1}-g}{2}\int_1^b dx\, \frac{1+J_0(\Lambda |z'|x)}{x}\biggr). 
\label{lebedev_variante}
\end{eqnarray}
Defining the new relative interaction
\begin{eqnarray}
\tilde v_{a'}(z',b)&:=bv_{a'}(z',b),
\end{eqnarray}
and keeping only linear terms in $\epsilon$ in Eq.~(\ref{lebedev_variante}), one obtains
\begin{eqnarray}
\frac{\tilde v_{a'}(z',b)-\tilde v_{a'}(z',1)}{\epsilon} &= \tilde v_{a'}(z',1)\Bigl[1-g^{-1}  +\frac{g^{-1}-g}{2}\bigl(1+J_0(\Lambda |z'|)\bigr)\Bigr].
 \label{renorm_step}
\end{eqnarray}
By defining a fictitious time variable  $t=\log b$ which runs from $0$ to $\infty$, the evolution equation becomes
\begin{eqnarray}
 \frac{1}{\tilde v_{a'}(z',t)}\frac{d\tilde v_{a'}}{dt}(z',t)= 1-g^{-1}+\frac{g^{-1}-g}{2}\bigl(1+J_0(\Lambda |z'|)\bigr).
\end{eqnarray}
With the initial condition $\tilde  v_{a'}(z',t=0)=v_{a'}(z')$, we get:
\begin{eqnarray}
\tilde v_{a'}(z',t)&=v_{a'}(z')  \exp\left(t\left\{1-g^{-1}+\frac{g^{-1}-g}{2}\left[1+J_0(\Lambda |z'|)\right] \right\} \right).
\label{solution}
\end{eqnarray}
One recovers the result that for non-interacting fermions ($g=1$), the relative XBH amplitude is not renormalized, as expected.
When the relative XBH  amplitude is short ranged, which amounts to taking $z'=0$, one finds $\tilde v=v e^{(1-g)\ln b}=vb^{1-g}$ which leads back to the localized impurity case of~\cite{kane_92a,kane_92b,kane_92c}. 

%%%%%%%%%%%%%%%%%%%%%%%%%%%%%%%%%%%%%%%%%%%%%%%%%%%%%%%%%%%%%%%%%

\section{Averages of exponentials}
\label{average_exponentials}

In this section, one computes the average of exponentials (over the free action) of linear combinations of $\theta$ and $\phi$ fields in terms of the four Green's functions in space and time. This is particularly relevant in the perturbative RG treatment of the relative XBH  amplitude $v_a(z)$ in Sec.~\ref{momentum_shell} and \ref{detailed_RG}, but also in the perturbative calculation of the current of Sec. \ref{perturbative_current}.

\subsection{Gaussian integrals}

Defining ${\vec{p}}=(q,\omega)$, we wish to compute the following average:
\begin{eqnarray}
{\cal I}&=\biggl\langle \exp\biggl( \pm\sum_{\vec{p}}\Bigr[ A_{\vec{p}}\phi_{\vec{p}}+A^*_{\vec{p}}\phi^*_{\vec{p}}+B_{\vec{p}}\theta_{\vec{p}}+B^*_{\vec{p}}\theta^*_{\vec{p}}\Bigr]\biggr)
\biggr\rangle.
\label{exp_average}
\end{eqnarray}
We use the notation $A^*_{\vec{p}}=A_{-\vec{p}}$ by analogy to  $\phi^*_{\vec{p}}=\phi_{-\vec{p}}$. However, while  
$\phi^*$ in the complex conjugate of  $\phi$, no such requirement exists for $A^*$ and $A$ (nor for  $B^*$ and $B$). In the following, we use a shorthand notation without the indices $\vec p$.
The average of an operator  ${\cal O}$ reads
\begin{eqnarray}
\bigl\langle {\cal O}\bigr\rangle=Z_0^{-1}\int{\cal D}\phi\, {\cal D}\theta\, e^{-S_0(\phi,\theta)}\cal{O},
\end{eqnarray}
where $S_0(\phi,\theta)$ is the bare Euclidean action of the LL wire Eq.~(\ref{bare_action_fourier}). Its shorthand notation reads
\begin{eqnarray}
S_0=\sum \bigl[ a\phi^*\phi+b\theta^*\theta+c(\phi^*\theta+\theta^*\phi)\bigl],
\end{eqnarray}
where $a=\frac{q^2g}{2}$, $b=\frac{q^2g^{-1}}{2}$ and $c=-\frac{i\omega q}{2}$.

Performing linear transformations of the fields $\theta$, $\phi$, and computing the relevant Gaussian integrals, one obtains:
\begin{eqnarray}
{\cal I}&=\exp \Bigl(-\sum \Bigl[AA^*a^{-1}+\frac{(B-Aca^{-1})(B^*-A^*ca^{-1})}{b-c^2a^{-1}}\Bigr]\Bigr).
\label{forgen}
\end{eqnarray}

%%%%%%%%%%%%%%%%%%%%%%%%%%%%%%%%%%%%%%%%%%%%%%%%

\subsection{Applications}
The two following averages turn out to be relevant for our study
\begin{eqnarray}
	{\cal I}_s&=\Bigl\langle \sin\Bigl(\sqrt{\pi}\bigl[ \theta(x+z/2,\tau)+\theta(x-z/2,\tau)\Bigr.\Bigr.\bigr.\nonumber\\	
&\Bigl.\Bigl.\Bigl.\qquad	+\phi(x+z/2,\tau)-\phi(x-z/2,\tau)\bigr] \Bigr)\Bigr\rangle\\
	{\cal I}_c&=\Bigl\langle \cos\Bigl(\sqrt{\pi}\bigl[ \theta(x+z/2,\tau)+\theta(x-z/2,\tau)\Bigr.\Bigr.\bigr.\nonumber\\	
&\Bigl.\Bigl.\Bigl.\qquad	+\phi(x+z/2,\tau)-\phi(x-z/2,\tau)\bigr] \Bigr)\Bigr\rangle.
\end{eqnarray}
We identify for this case the coefficients  $A$, $A^*$, $B$, $B^*$ by using the Fourier transforms of the fields. One thus obtains the following relations:
\begin{eqnarray}
	i\sqrt{\pi}\bigl(\phi(x+z/2,\tau)-\phi(x-z/2,\tau)\bigr)&=
	\sum_{\vec p} A_{\vec p}\phi_{\vec p}+A^*_{\vec p}\phi^*_{\vec p}\\
	i\sqrt{\pi}\bigl(\theta(x+z/2,\tau)+\theta(x-z/2,\tau)\bigr)&=
\sum_{\vec{p}}B_{\vec{p}}\theta_{\vec{p}}+B^*_{\vec{p}}\theta^*_{\vec{p}},
\end{eqnarray}
with  $A_{\vec{p}}=A^*_{-\vec p}=-\sqrt{\pi}\sin(qz/2)e^{i(qx+\omega \tau)}$ and $B_{\vec{p}}=B^*_{-\vec p}=i\sqrt{\pi}\cos(qz/2)e^{i(qx+\omega \tau)}$.

One can then apply the Gaussian integral result. In the case of a sine:
\begin{eqnarray}
	{\cal I}_s
	&=&0,
\end{eqnarray}
because of the symmetry properties of the exponent under sign reversal of $A$ and $B$. For a cosine, we obtain:
\begin{eqnarray}
	{\cal I}_c
	&=&\biggl\langle \exp\biggl( \sum_{\vec{p}}\Bigl[A_{\vec{p}}\phi_{\vec{p}}+A_{-\vec{p}}\phi^*_{\vec{p}}+B_{\vec{p}}\theta_{\vec{p}}+B_{-\vec{p}}\theta^*_{\vec{p}}\Bigr]\biggr)\biggr\rangle,
\end{eqnarray}
which has precisely the form of Eq.~(\ref{exp_average}). In order to apply the result of Eq.~(\ref{forgen}), we need the following identities:
\begin{eqnarray}
	b-c^2a^{-1}&=&\frac{q^2+\omega^2}{2g}=\frac{{\vec{p}}^2}{2g},\\
	A_{\vec{p}}A_{-\vec{p}}a^{-1}&=&-\frac{2\pi}{q^2g}\sin^2(qz/2),
\end{eqnarray}
\begin{eqnarray}
&B_{\vec{p}}B_{-\vec{p}}+A_{\vec{p}}A_{-\vec{p}}c^2a^{-2}-B_{-\vec{p}}A_{\vec{p}}ca^{-1}-B_{\vec{p}}A_{-\vec{p}}ca^{-1}&\nonumber\\
	& \qquad =-\pi\cos^2(qz/2)+\frac{\pi\omega^2}{q^2g^2}\sin^2(qz/2).&
\end{eqnarray}
Using the Fourier transform version of the Green's functions Eq.~(\ref{fourier_green}), 
we obtain the argument of the exponential in terms of space and imaginary time Green's functions:
\begin{eqnarray}
	&\sum AA^*a^{-1}+(B-Aca^{-1})(B^*-A^*ca^{-1})(b-c^2a^{-1})^{-1} \nonumber \\
	&\qquad=\-\pi\Bigl[ G^{\theta\theta}(0,0)+G^{\phi\phi}(0,0)+G^{\theta\theta}(z,0)-G^{\phi\phi}(z,0)\Bigr].
\end{eqnarray}
With Eq.~(\ref{forgen}), we conclude that:
\begin{eqnarray}
{\cal I}_c&=e^{-\pi\bigl[ G^{\theta\theta}(0,0)+G^{\phi\phi}(0,0)+G^{\theta\theta}(z,0)-G^{\phi\phi}(z,0)\bigr]}.
\label{cos_integral}
\end{eqnarray}
This is precisely the functional integral that we need to compute when integrating the fast degrees of freedom in the perturbative RG procedure in Sec. \ref{momentum_shell}: there, $G^{\theta\theta}$, $G^{\phi\phi}$ are replaced by their fast versions $\tilde G^{\theta\theta}$, $\tilde G^{\phi\phi}$.  This is also the same functional integral that we need for computing $Z_1$ in Sec. \ref{perturbative_current}. 

%%%%%%%%%%%%%%%%%%%%%%%%%%%%%%%%%%%%%%%%%%%%%%%%%

\section{Perturbative calculation of the partition function}
\label{perturbative_appendix}

We first show that $Z_1=0$. As shown in \ref{average_exponentials}, the argument of the exponential contains both sums and differences of Green's functions in imaginary time and space as is obvious from Eqs.~(\ref{Z1}) and (\ref{cos_integral}).  

Let $\alpha$ denote the large momentum cutoff [$\alpha \propto \eta^{-1}$, with $\eta$ the short wave length cutoff introduced in Eq.~(\ref{eq:psi})] and $\epsilon$ denote an infrared cutoff. We use the notation $\vec {x}=(z,\tau)$. From~\cite{abramowitz}, after regularization of the integrals defining the Green's functions, one finds:
\begin{eqnarray}
G^{\theta\theta}(\vec x)=
\left\{
\begin{array}{ll}
 \frac{g}{2\pi} (\ln\alpha-\ln\epsilon) &\rm{if~} |\vec x|=0 \\ 
 \frac{g}{2\pi} \biggl[-\ln\frac{\epsilon |\vec x|}{2}-\gamma+\int\limits_0^{\epsilon|\vec x|}dp\,\frac{1-J_0(p)}{p}\biggr] & \rm{if~} |\vec x|\neq 0.
\end{array}
\right.
\end{eqnarray}
Therefore the difference between two Green's functions have the form: 
\begin{eqnarray}
G^{\phi\phi}(0,0)-G^{\phi\phi}(0,0)&:=0\\
G^{\phi\phi}(0,0)-G^{\phi\phi}(\vec x)&=\frac{g^{-1}}{2\pi}\left[ \ln \alpha+\ln \frac{|\vec x|}{2}+ \gamma\right]\label{infinite}\\
G^{\phi\phi}(\vec x)-G^{\phi\phi}(\vec y)&=\frac{g^{-1}}{2\pi}\ln\frac{|\vec y|}{|\vec x|},
\end{eqnarray}
where $|\vec x|$ and $|\vec y|$ are non-zero.
The same results are obtained for $G^{\theta\theta}$ by operating the substitution $g\to g^{-1}$. This gives a finite result when $\vec x=\vec y=\vec 0$, when $|\vec x|\neq 0$ and $|\vec y|\neq 0$, and a positive infinite result for the case of Eq.~(\ref{infinite}) if we let $\alpha$ be infinite. In the following, we keep the cutoff dependent expression of Eq.~(\ref{infinite}).

On the other hand the sum of two Green's functions read: 
\begin{eqnarray}
G^{\theta\theta}(0,0)+G^{\theta\theta}(0,0)&=\frac{2g}{2\pi}(\ln\alpha-\ln\epsilon)\\
G^{\theta\theta}(0,0)+G^{\theta\theta}(\vec x)&=\frac{g}{2\pi}\biggl[\ln \alpha -2\ln \epsilon-\ln\frac{|\vec x|}{2}-\gamma\nonumber\\
&\quad\Bigl.+\int_0^{\epsilon|\vec x|}dp\,\frac{1-J_0(p)}{p}\biggr]\\
G^{\theta\theta}(\vec x)+G^{\theta\theta}(\vec y)&=\frac{g}{2\pi}\biggl[-2\ln\epsilon-\ln\frac{|\vec x||\vec y|}{4}-2\gamma\nonumber\\
&\quad+\int_0^{\epsilon|\vec x|}dp\ \frac{1-J_0(p)}{p}+\int_0^{\epsilon|\vec y|}dp\ \frac{1-J_0(p)}{p}\biggr],
\end{eqnarray}
where $|\vec x|$ and $|\vec y|$ are non-zero.
We conclude that all sums of Green's functions (regardless of zero or non-zero arguments) give a positive infinite result as the infrared cutoff $\epsilon$ tends to zero. $Z_1$ is therefore the exponential of a negative infinite constant, so that $Z_1=0$.

The calculation of $Z_2$ involves the average of the product of two exponentials, see Eq.~(\ref{Z2}). We thus study the following average: 
\begin{eqnarray}
\mathcal J_\sigma&=\left\langle e^{i\sqrt\pi\bigl[\theta(1,\tau)+\theta(2,\tau)+\phi(1,\tau)-\phi(2,\tau)\bigr]}  e^{i \sigma \sqrt\pi\bigl[\theta(1',\tau')+\theta(2',\tau')+\phi(1',\tau')-\phi(2',\tau')\bigr]} \right\rangle,
\label{j_sigma}
\end{eqnarray} 
where we use the shorthand notations $\delta\tau = \tau-\tau'$, $``1"=x+z/2$, $``2"=x-z/2$,  $``1'"=x'+z'/2$, $``2'"=x'-z'/2$ and $\sigma=\pm$. Further defining 
$``kx"=qx+\omega\tau$, $``kx'"=qx'+\omega\tau'$, $s=\sin qz/2$, $s'=\sin qz'/2$, $c=\cos qz/2$ and $c'=\cos qz'/2$. 

We then notice that the expression for $\mathcal J_\sigma$ in Eq.~(\ref{j_sigma}) has the same form as the average Eq.~(\ref{exp_average}) with the redefinition of the parameters:
\begin{eqnarray}
A_{\vec p}&=-\sqrt\pi se^{ikx}-\sigma \sqrt\pi s'e^{ikx'}\\
A_{-\vec p}&= \sqrt\pi se^{-ikx}+\sigma\sqrt\pi s'e^{-ikx'}\\
B_{\vec p}&=i\sqrt\pi ce^{ikx}+\sigma i\sqrt\pi c'e^{ikx'}\\
B_{-\vec p}&=i\sqrt\pi ce^{-ikx}+\sigma i\sqrt\pi c'e^{-ikx'}.
\end{eqnarray}
After some tedious algebra we thus get the result:
\begin{eqnarray}
\mathcal J_\sigma&= \exp\Bigl( -\pi\Bigl[G^{\theta\theta}(\vec 0)+G^{\phi\phi}(\vec 0)+G^{\theta\theta}(z,0)-G^{\phi\phi}(z,0) \nonumber \\
&+G^{\theta\theta}(\vec 0)+G^{\phi\phi}(\vec 0)+G^{\theta\theta}(z',0)-G^{\phi\phi}(z',0) \nonumber \\
&+\sigma\bigl[G^{\theta\theta}(1-1',\delta \tau)+G^{\phi\phi}(1-1',\delta \tau)+G^{\theta\theta}(1-2',\delta \tau)-G^{\phi\phi}(1-2',\delta \tau)\bigr] \nonumber \\
&+\sigma\bigl[G^{\theta\theta}(2-2',\delta \tau)+G^{\phi\phi}(2-2',\delta \tau)+G^{\theta\theta}(2-1',\delta \tau)-G^{\phi\phi}(2-1',\delta \tau)\bigr] \nonumber \\
& +2\sigma \bigl[ G^{\theta\phi}(1-1',\delta \tau)-G^{\theta\phi}(2-2',\delta \tau)\bigr]\Bigr]\Bigr).
\end{eqnarray}
We now make the following statements. First, the exponential of mixed Green functions $G^{\theta\phi}$ corresponds to a phase factor since such Green's functions are specified as follows: 
\begin{eqnarray}
 G^{\theta\phi}(x,\tau)&=\int\frac{d^2p}{(2\pi)^2}\frac{i\omega/q}{\omega^2+q^2}e^{iqx+i\omega \tau} \nonumber \\
&= \frac{-i}{2\pi}\arctan(x/\tau)\bigl[\Theta(\tau)+\Theta(-\tau)\bigr],
\end{eqnarray}
where $\Theta$ is the Heaviside function. Second, using the previous results concerning the sum and difference of Green's functions $G^{\theta\theta}$ (or $G^{\phi\phi}$), one can show that $\mathcal J_+=0$. 
Indeed, with the above result concerning mixed Green's functions, we see that the modulus of $\mathcal J_+$ equals zero (it is the exponential of a negative infinite constant):
\begin{eqnarray}
|\mathcal J_+|&= \exp\Bigl( -\pi\Bigl[G^{\theta\theta}(\vec 0)+G^{\phi\phi}(\vec 0)+G^{\theta\theta}(z,0)-G^{\phi\phi}(z,0) \nonumber \\
&+G^{\theta\theta}(\vec 0)+G^{\phi\phi}(\vec 0)+G^{\theta\theta}(z',0)-G^{\phi\phi}(z',0) \nonumber \\
&+G^{\theta\theta}(1-1',\delta \tau)+G^{\phi\phi}(1-1',\delta \tau)+G^{\theta\theta}(1-2',\delta \tau) \nonumber \\
&-G^{\phi\phi}(1-2',\delta \tau)+G^{\theta\theta}(2-2',\delta \tau)+G^{\phi\phi}(2-2',\delta \tau) \nonumber \\
&+G^{\theta\theta}(2-1',\delta \tau)-G^{\phi\phi}(2-1',\delta \tau)\Bigr]\Bigr) \nonumber \\
&=0.
\end{eqnarray} 
We conclude that the only non-vanishing contribution to $Z_2$ in Eq.~(\ref{Z2}) is $\mathcal J_-$:
\begin{eqnarray}
\mathcal J_-&= \exp\Bigl( -\pi\Bigl[G^{\theta\theta}(\vec 0)+G^{\phi\phi}(\vec 0)+G^{\theta\theta}(z,0)-G^{\phi\phi}(z,0) \nonumber \\
&+G^{\theta\theta}(\vec 0)+G^{\phi\phi}(\vec 0)+G^{\theta\theta}(z',0)-G^{\phi\phi}(z',0) \nonumber \\
&-G^{\theta\theta}(1-1',\delta \tau)-G^{\phi\phi}(1-1',\delta \tau) -G^{\theta\theta}(1-2',\delta \tau) \nonumber \\
&+G^{\phi\phi}(1-2',\delta \tau)-G^{\theta\theta}(2-2',\delta \tau)-G^{\phi\phi}(2-2',\delta \tau)  \nonumber \\
&-G^{\theta\theta}(2-1',\delta \tau)+G^{\phi\phi}(2-1',\delta \tau)  \nonumber \\
&-2 G^{\theta\phi}(1-1',\delta \tau)+2G^{\theta\phi}(2-2',\delta \tau)\Bigr]\Bigr).
\label{j-final}
\end{eqnarray}
It is expressed as the exponential of differences of diagonal Green's functions $G^{\theta\theta}$ and $G^{\phi\phi}$, along with a phase factor [the last line of Eq.~(\ref{j-final})]. This completes the details of the perturbative analysis. 
 
%%%%%%%%%%%%%%%%%%%%%%%%%%%%%%%%%%%%%%%%%%%%%%%%%

\section{Smooth cutoff functions}
\label{softcutoff_appendix}

\begin{figure}[htbp]
	\centering
		\includegraphics[width=1\textwidth]{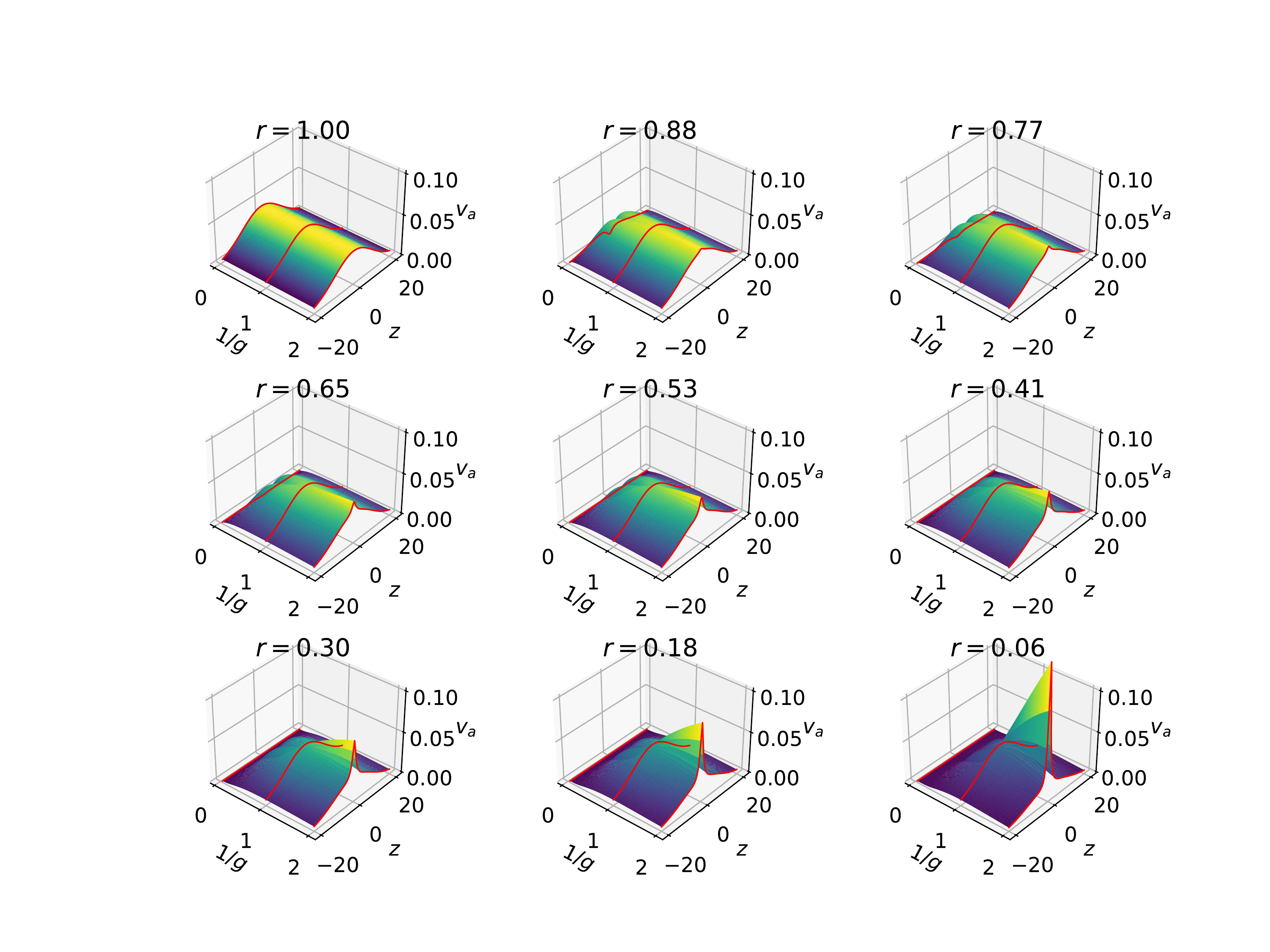}
		\caption{Three-dimensional evolution of the non-local part of a relative Gaussian backscattering interaction $v_a(z)$ with typical width $a=10$ (all length scales in units of $\Lambda^{-1}$) as a function of the relative coordinate $z$ and the inverse LL parameter $1/g$. The quantity $r$ on each graph represents the ratio of the running cutoff to the initial cutoff. Here, a smooth cutoff function is used which removes the oscillation of the Bessel function $J_0$. Qualitatively, the same phenomenon occurs: if $1/g<1$ the RG flow suppresses the interaction; if $1/g>1$ the interaction evolves as a Dirac-delta like function and the line at $1/g=1$ is kept approximately invariant with $1\leq e^{\alpha(g=1,\Lambda,z)t}<1.00026$ for all $t$ that are considered here.}
		\label{smooth_cutoff_function}
\end{figure}

The oscillations in the non-local part of a relative Gaussian backscattering interaction $v_a(z)$ is due to the hard cutoff in the integrals defining the Green's functions. We can introduce smooth cutoff functions $f$ where we basically want $0\leq f(p;\Lambda)\leq 1$ for all $p$, $f(p;\Lambda)=1$ if $0\leq p\ll \Lambda$ and $f(p;\Lambda)=0$ if $p\gg \Lambda$. Then we make the substitution $\int_0^\Lambda dp\to \int_0^\infty dp\ f(p;\Lambda)$ so that, for example, for $b=1+\epsilon$ with $0<\epsilon\ll 1$, the Green's function $\tilde G^{\theta''\theta''}(z',0)$ now reads
\begin{eqnarray}
\tilde G^{\theta''\theta''}(z',0)&=\frac{g}{2\pi}\int_\Lambda^{b\Lambda} dp\  \frac{J_0(p|z'|)}{p}   \nonumber \\
&=\frac{g}{2\pi}\int_0^\infty dp\ \Bigl[ f(p;b\Lambda)-f(p;\Lambda)\Bigr]\frac{J_0(p|z'|)}{p}   \nonumber \\
&= \epsilon \frac{g}{2\pi}\int_0^\infty dp\ \Bigl.\partial_b f(p;b\Lambda)\Bigr|_{b=1}\frac{J_0(p|z'|)}{p}.
\end{eqnarray}
Following the same procedure as before, we obtain
\begin{eqnarray}
\alpha(g,\Lambda,z) &= 1- \frac{g+g^{-1}}{2}\int_0^\infty dp\ \frac{\Bigl.\partial_b f(p;b\Lambda)\Bigr|_{b=1}}{p}   \nonumber \\
&-\frac{g-g^{-1}}{2}\int_0^\infty dp\ \Bigl.\partial_b f(p;b\Lambda)\Bigr|_{b=1}\frac{J_0(p|z'|)}{p}.
\end{eqnarray}
For example, the following cutoff function $f(p;\Lambda)=\Lambda^2/(p^2+\Lambda^2)$ is used in the RG procedure as shown in Fig.~\ref{smooth_cutoff_function}. Another possibility is to use $f(p;\Lambda)=\exp\Bigl(1-1/(1-p^2/\Lambda^2)\Bigr)$ if $|p|<\Lambda$ and $0$ if $|p|\geq \Lambda$. This cutoff function leads to the 
same qualitative behavior (not shown): the quasi-suppression of the oscillations.

%%%%%%%%%%%%%%%%%%%

\bibliography{popoff_2}

\end{document}